\documentclass[12pt,preprint]{emulateapj}

\usepackage{amssymb}
\usepackage{times}
\usepackage{amsmath}
\usepackage{subfigure}
\usepackage{multirow}
\usepackage[colorlinks,backref,dvipdfm,linkcolor=blue,anchorcolor=,citecolor=blue]{hyperref}
\makeatletter

\newcommand{\rmnum}[1]{\romannumeral #1}
\newcommand{\Rmnum}[1]{\expandafter\@slowromancap\romannumeral #1@}
\makeatother

\slugcomment{}

\shorttitle{\textit{A new C-D-like diagram for SPB stars: $f_{\Delta P}$ vs. $\langle\Delta P\rangle$}}
\shortauthors{\textit{Wu, Li \& Deng}} 

\begin{document}


\title{A new C-D-like diagram for SPB stars: The variations of period spacing as a signature of evolutionary status}


\author{
Tao Wu\altaffilmark{1,2,3}, Yan Li \altaffilmark{1,2,3,4} and Zhen-min Deng\altaffilmark{1,2,4}
}


\altaffiltext{1}{Yunnan Observatories, Chinese Academy of Sciences, 396 Yangfangwang, Guandu District, Kunming, 650216, P. R. China; wutao@ynao.ac.cn, ly@ynao.ac.cn}
\altaffiltext{2}{Key Laboratory for the Structure and Evolution of Celestial Objects, Chinese Academy of Sciences, 396 Yangfangwang, Guandu District, Kunming, 650216, P. R. China}
\altaffiltext{3}{Center for Astronomical Mega-Science, Chinese Academy of Sciences, 20A Datun Road, Chaoyang District, Beijing, 100012, P. R. China}
\altaffiltext{4}{University of Chinese Academy of Sciences, Beijing 100049, China}


\begin{abstract}
  The Slowly Pulsating B-type (SPB) stars are the upper main-sequence stars on the HR diagram. Their oscillations are high-order, low-degree g-mode and can be used to probe the structure of the radiative zone where is near the outer boundary of the convective core and constrain the chemical mixing in stellar interiors. In SPB stars, the period spacing periodically varies with periods. It has regarded as a signature of the chemical composition gradient beyond the convective core. Based on theoretical calculations, we find that the variation frequency of the period spacings ($f_{\Delta P}$) is related to the width of the $\mu$-gradient region on the buoyancy radius ($\Lambda_{\mu}$) with the relation of $f_{\Delta P}\sim0.5\Lambda_{\mu}$. This indicates that the variation frequency $f_{\Delta P}$ is sensitive to the central hydrogen mass fraction $X_{\rm C}$ (i.e., the evolutionary status). Finally, we find that the variation frequency $f_{\Delta P}$ and the means of the period spacings $\langle\Delta P\rangle$ can be used to construct a new C-D-like diagram ($f_{\Delta P}$ vs. $\langle\Delta P\rangle$) which can be used to roughly decide the stellar evolutionary stages and to approximately determine stellar mass for SPB stars.
\end{abstract}



\section{Introduction}\label{sec-intr}

Theoretically, stellar oscillation modes are roughly divided into two groups according to the difference of the restoring forces. They are pressure modes (p-mode) with the restoring forces of pressure and gravity modes (g-mode) with the restoring force of buoyancy, respectively. Therefore, the g-mode merely stably propagates in the radiative zones in stellar interiors. And the oscillation power spectra shows almost uniformly-spaced in periods if the star has homogeneous composition, i.e., the period spacing ($\Delta P_{n,l}=P_{n+1,l}-P_{n,l}$) is almost a constant.

In fact, the stratification of the chemical composition in stellar interiors almost appears through all of the main-sequence and post-main-sequence stars, such as main sequence stars, red giants, white dwarfs. The steep composition gradient ($\mu$-gradient) produces a steep gradient in the density which will reflect the propagation of the oscillation waves as a reflecting boundary and leads to the phenomenon of the ``mode trapping". This has been interpreted as the signature of chemical composition gradients in stellar interiors \citep[see e.g.,][]{Miglio2008MNRAS}. Due to mode trapping, the pulsating mode will deviate from the uniform period spacing which is the result of the first-order approximation of g-mode oscillations (see Equation \eqref{eq:Dp} of Section \ref{sec.dp}). They have been theoretically investigated, observed, and applied in different pulsating stars, such as pulsating white dwarfs \citep[see e.g.,][]{Winget1981ApJ,Kawaler1995ASPC,Metcalfe2003MNRAS}, main sequence stars: $\gamma$ Doradus and Slowly Pulsating B (SPB) stars \citep[see e.g.,][]{Miglio2008MNRAS,Moravveji2015AA}, and red giants \citep[see e.g.,][]{Jiang2014MNRAS}.

SPB stars are the upper main-sequence stars of intermediate mass ($2.5\sim8~M_{\odot}$) \citep[see e.g.,][]{Aerts2010}. It is therefore of a convective core and a radiative envelope. Its convective core decreases with its evolution and leaves a chemical composition gradient region beyond the convective core. Finally, such chemical composition gradient brings a periodical signal on the oscillating periods.

The effective temperature of SPB stars ranges from 11000 K to 22000 K. The period of non-radial, high-order, low-degree g-modes ranges from about 0.5 to 3 days \citep[more review refer to][Sec. 2.3.6 (p60)]{Aerts2010}. Recently, many SPB stars have been observed by {\it CoRoT} \citep[e.g.,][]{Baglin2006cosp}, {\it Kepler} \citep[e.g.,][]{Borucki2010Sci,Koch2010ApJ,Gilliland2010PASP}, and {\it K2} \citep[e.g.,][]{Haas2014AAS,Howell2014PASP} telescopes \citep[e.g.,][]{Degroote2012AA,Degroote2010Natur,Papics2014AA,Papics2017AA,Moravveji2015AA,Moravveji2016ApJ}. The works of \citet[][]{Papics2014AA,Papics2017AA}, \citet[][]{Moravveji2015AA,Moravveji2016ApJ}, and \citet[][]{Triana2015ApJ} have shown that those stars have larger convective cores during hydrogen burning in the stellar center and many of them are faster rotators. Their oscillation period spacings are almost quasi-equal \citep[e.g.,][]{Degroote2010Natur,Papics2012AA,Papics2014AA} but with small deviations if those modes are not split by rotation and/or other physical processes, such as magnetic field. The deviation of quasi-equal period spacing carries information about the chemical mixing in stellar interiors \citep[see e.g.,][]{Miglio2008MNRAS,Degroote2010Natur,Moravveji2015AA}. They have been used to constrain the evolutionary status, overshooting extent, and extra mixing in SPB stars by, for instance, \citet[][HD 50230]{Degroote2010Natur}, \citet[][KIC 10526294]{Moravveji2015AA}, and \citet[][KIC 8324482]{Deng2018}.

The previous works suggested that the high-order g-mode of SPB stars can be used to efficiently probe the interior structure and status, such as the size of convective core, the shape of chemical elements in convective core overshooting region, and the mass fraction of hydrogen in stellar center. \citet[][]{Miglio2008MNRAS} has been made a series of theoretical analyzing for SPB stars and $\gamma$ Doradus and found that the variation of the period spacings are dependent on a mount of physical processes , such as overshooting, diffusion, and rotations. In other words, it can be used to constrain those physical processes. Inspired by the previous works and the detail analysis of \citet[][]{Miglio2008MNRAS} for the variation of period spacings, we further study the properties of the period spacings of high-order, low-degree ($l=1,~m=0$) g-modes and try to construct a new diagram, which is similar to the C-D diagram of \citet{jcd1984srps.conf} and can be used to constrain the stellar evolutionary stages for SPB stars.

\section{Period spacing periodically varies with the period}\label{sec.dp}
It is well known that the period spectra of the g-mode is fully determined by the spatial distribution of the buoyancy frequency (as-called Brunt-V\"{a}is\"{a}l\"{a} frequency; $N$) which is defined as \citep[see e.g.,][]{Aerts2010}
\begin{equation*}\label{eq:N1}
   N^2=g\left(\frac{1}{\Gamma_{1}p}\frac{{\rm d}p}{{\rm d}r}-\frac{1}{\rho}\frac{{\rm d}\rho}{{\rm d}r}\right).
\end{equation*}
For fully ionized ideal gas, buoyancy frequency $N$ can be approximately expressed as
\begin{equation}\label{eq:N2}
   N^2\simeq\frac{g^2\rho}{p}(\nabla_{\rm ad}-\nabla{\mathbf{+}}\nabla_{\mu}),
\end{equation}
where, $\nabla_{\rm ad}$, $\nabla$, and $\nabla_{\mu}$ are the adiabatic temperature gradient, the temperature gradient, and $\mu$-gradient, respectively.  Based on the first-order asymptotic approximation of \citet[][]{Tassoul1980ApJS}, the period spacings of low-degree, high-order g modes can be expressed as
\begin{equation}\label{eq:Dp}
  \Delta \Pi_{l}=\frac{2\pi^2}{L}\left(\int^{R}_{0}\frac{N}{r}{\rm d}r\right)^{-1},
\end{equation}
where $L=\sqrt{l(l+1)}$. Correspondingly, the period can be expressed as $P_{n,l}=\Delta\Pi_{l}(n+\phi)$, where $\phi$ is the phase offset of the pulsating mode.

Similar to the definition of the acoustic depth ($\tau(r)=\int^{R}_{r}\frac{{\rm d}r'}{c_{\rm s}}$ with the unit of time, i.e., second or day; acoustic radius $\tau_{0}=\tau(0)$), the buoyancy depth is defined as
\begin{equation}\label{eq:Lambda}
   \Lambda(r)=\int^{R}_{r}\frac{N}{2\pi}\frac{{\rm d}r'}{r'},
\end{equation}
with the unit of frequency, i.e., $\mu$Hz or Hz (more description for buoyancy depth and radius see APPENDIX A of \citet[][]{Wu2018}). Obviously, period spacing $\Delta \Pi_{l}$ and buoyancy radius $\Lambda_{0}$ ($=\Lambda(0)$) are related to each other through the relation of:
\begin{equation}\label{eq:DP-L0}
\Delta \Pi_{l}=\frac{\pi}{L}\Lambda_{0}^{-1}.
\end{equation}

\begin{figure}
  \begin{center}
\includegraphics[scale=0.35,angle=-90]{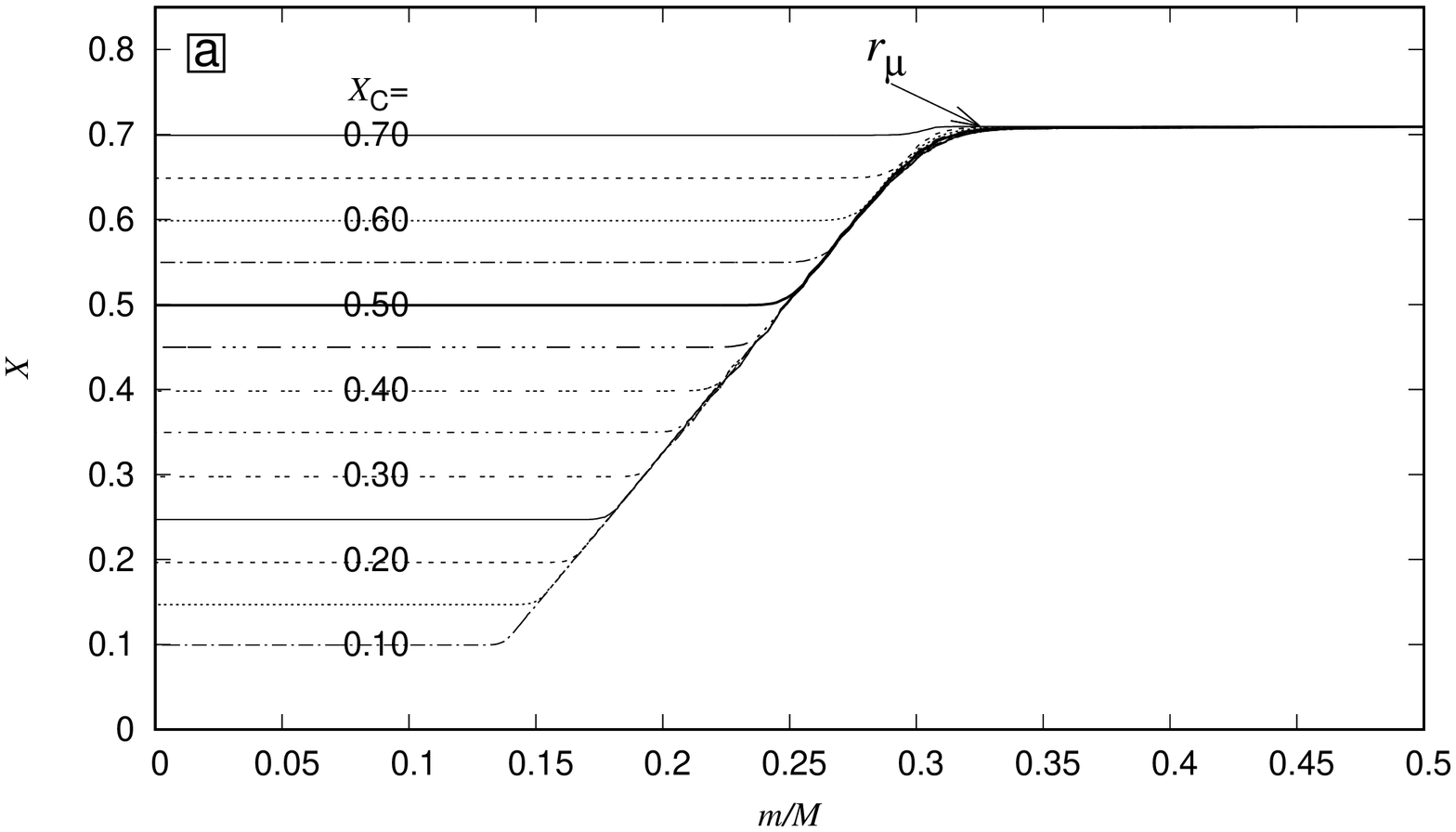}
\includegraphics[scale=0.35,angle=-90]{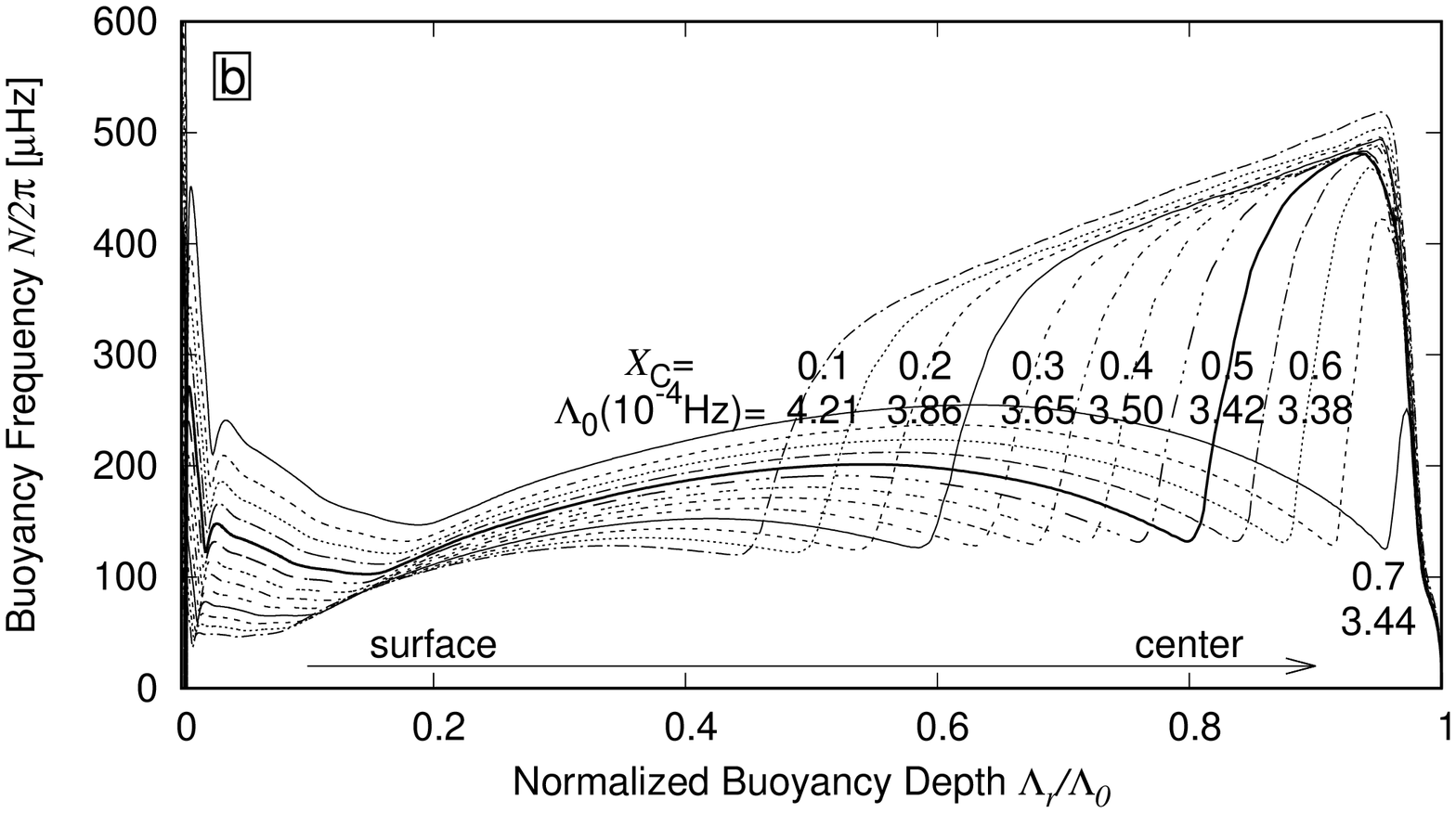}
\includegraphics[scale=0.35,angle=-90]{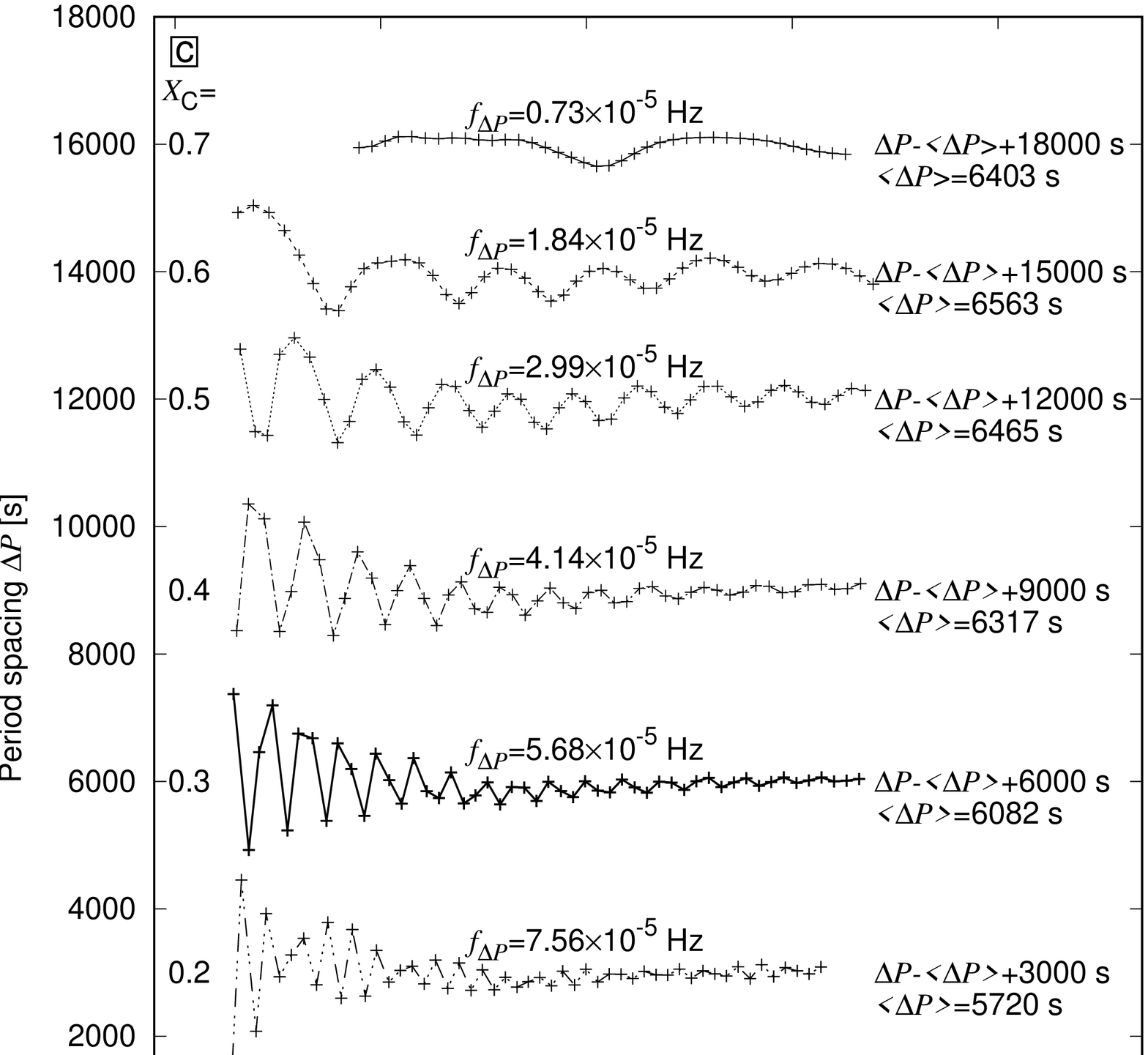}
   \caption{{\bf (a):} The hydrogen mass fraction of stellar interiors ($X$) as a function of stellar mass ($m/M$); {\bf (b):} The Buoyancy frequency ($N/2\pi$) as a function of normalized buoyancy depth ($\Lambda_{r}/\Lambda_{0}$); {\bf (c):} Period spacings ($\Delta P$) as a function of periods ($P$) of $l=1$, $m=0$ modes. In panels, the different kinds of line types represent the different evolutionary stages. Center hydrogen mass fraction ($X_{\rm C}$) drops from 0.7 to 0.1 with a step of 0.05 for Panels (a) and (b) and with a step of 0.1 for Panel (c), respectively. The initial inputs of those models are $M=4.0~\rm M_{\odot}$, $Z=0.014$, $X=0.71$, $f_{\rm ov}=0.020$, and $\log D_{\rm mix}=2.0$. For parts of models, their $\Lambda_{0}$, $f_{\Delta P}$, and $\langle{\Delta P}\rangle$ are marked with text in figure.
  }\label{fig.1}
  \end{center}
\end{figure}

Correspondingly, in the case of a model consisting of an inner convective core and an outer radiative envelope, for instance the SPB stars, the width of the $\mu$-gradient regions in buoyancy size is defined as
\begin{equation}\label{eq:Lambdamu}
   \Lambda_{\mu}=\int^{r_{\mu}}_{0}\frac{N}{2\pi}\frac{{\rm d}r'}{r'}.
\end{equation}
where $r_{\mu}$ denote the position of the outer boundary of the $\mu$-gradient region (see Figure \ref{fig.1}(a)). In Figure \ref{fig.1}, panels (a) and (b) represent the profile of hydrogen mass fraction ($X$) and the buoyancy frequency $N$ against mass ($m/M$) and normalized buoyancy radius ($\Lambda_{r}/\Lambda_{0}$) for different evolutionary status models, respectively. In addition, panel (c) presents their period spacings ($\Delta P$) against periods ($P$). As shown in Figures \ref{fig.1}(a) and \ref{fig.1}(b), both $\Lambda_{0}$ and $\Lambda_{\mu}$ increase with the decreasing of the central hydrogen mass fraction $X_{\rm C}$ for a given evolutionary track.

Based on theoretical analysis of \citet[][]{Tassoul1980ApJS}, \citet[][]{Miglio2008MNRAS} reported that there is a sinusoidal component on period spacings for $\gamma$ Doradus and SPB stars. It is expressed as
\begin{equation}\label{eq:sindp}
\delta P_{n}\propto A_{\delta P}\cos( 2L\Lambda_{\mu}P_{n}+\frac{\pi}{2})
\end{equation}
where the amplitude $A_{\delta P}$ is related to the shape of the buoyancy frequency $N$ in the $\mu$-gradient region \citep[for more descriptions of the theoretical asumptions and deductions please refer to][]{Miglio2008MNRAS}.

For a step function ($\frac{\delta N}{N}=\frac{1-\alpha^2}{\alpha^2}H(r_{\mu}-r)$, $H(r_{\mu}-r)$ is the step function; see Equation (11) and left-hand panel of Figure 5 of \citet{Miglio2008MNRAS}), the amplitude is a constant and expressed as
\begin{equation}\label{eq:A1}
A_{\delta P}=\frac{1-\alpha^2}{\alpha^2}(2\pi L\Lambda_{0})^{-1},
\end{equation}
where $\alpha=\sqrt{N_{+}/N_{-}}$ ($N_{-}$ and $N_{+}$ are the values of buoyancy frequency at the outer and inner border of the $\mu$-gradient region). On the other hands, for a ramp function ($\frac{\delta N}{N}=\frac{1-\alpha^2}{\alpha^2}\frac{r_{\mu}-r}{r_{\mu}-r_{0}}H(r_{\mu}-r)$; see Equation (15) and right-hand panel of Figure 5 of \citet{Miglio2008MNRAS}), the amplitude $A_{\delta P}$ modulated by a factor of $1/P_{n}$, i.e., $A_{\delta P}$ decreases with the increasing period $P_{n}$ or the radial order $|n|$. It is expressed as
\begin{equation}\label{eq:Ap}
A_{\delta P}=\frac{1}{P_{n}}\frac{1-\alpha^2}{\alpha^2}\frac{1}{4\pi^4 L\Lambda_{0}\Lambda_{\mu}}.
\end{equation}
Both of the observations \citep[see e.g.,][]{Degroote2010Natur,Papics2014AA,Papics2015ApJL,Papics2017AA,Moravveji2015AA} and the theoretical models \citep[see Figure \ref{fig.1}(c) of the present work, as well as the work of][]{Miglio2008MNRAS} illustrate that the amplitude of the variation on period spacing $A_{\delta P}$ decreases with the increasing of period $P_{n}$.

\section{Physical inputs and model calculations}

In the present work, our theoretical models were computed by the Modules of Experiments in Stellar Astrophysics (MESA), which is developed by \citet{MESA2011}. It can be used to calculate both the stellar evolutionary models and their corresponding oscillation information \citep{MESA2013}. We adopt the package {\small \textbf{``pulse"}} of version {\bf \small ``v6208"} to make our calculations for both stellar evolutions and oscillations \citep[for more detailed descriptions refer to][]{MESA2011,MESA2013}. The package {\small \textbf{``pulse"}} is a test suite example of MESA in the directory of MESA/star/test\_suite/pulse. In the present work, the oscillations (i.e., the periods of $l=1$, $m=0$ modes) are calculated with ADIPLS code (the Aarhus adiabatic oscillation package), which is developed by \citet{jcd2008} and added in MESA.

Based on the default parameters, we adopt the OPAL opacity table GS98 \citep{gs98} series. We choose the Eddington grey-atmosphere $T-\tau$ relation as the stellar atmosphere model, and treat the convection zone by the standard mixing-length theory (MLT) of \citet{cox1968} with mixing-length parameter $\alpha_{\rm MLT}=2.0$.

We adopt the theory of \citet[][]{Herwig2000} to treat the convective overshooting of the convective core. The overshooting mixing diffusion coefficient $D_{\rm ov}$ exponentially decreases with distance which extends from the outer boundary of the convective core with the Schwarzschild criterion, i.e.,
\begin{equation}\label{eq_Dov}
D_{\rm ov}=D_{\rm conv,0}\exp{\left(-\frac{2z}{f_{\rm ov}H_{P,0}}\right)},
\end{equation}
where $D_{\rm conv,0}$ and $H_{P,0}$ are the MLT derived diffusion coefficient near the Schwarzschild boundary and the corresponding pressure scale height at that location, respectively. $z$ is the distance in the radiative layer away from that location. $f_{\rm ov}$ is an adjustable parameter \citep[for more detailed discriptions refer to][]{Herwig2000,MESA2011}.

In addition, the element diffusion, semi-convection, thermohaline mixing, and the mass-loss were not included in the theoretical models.

According to the theory of stellar structure and evolution, the convective overshooting in the convective core will generate a relatively larger convective core and leads to a larger period spacing ($\Delta\Pi$; see Equation \eqref{eq:Dp} in Section \ref{sec.dp}) for main sequence stars \citep[see e.g.,][]{Miglio2008MNRAS,Degroote2010Natur}. It is necessary to take the convective overshooting of convective core into account in theoretical models for the upper main-sequence stars \citep[see discussions in][for example]{Saio2014psce}.

\begin{figure*}
  \begin{center}
\includegraphics[scale=0.45,angle=-90]{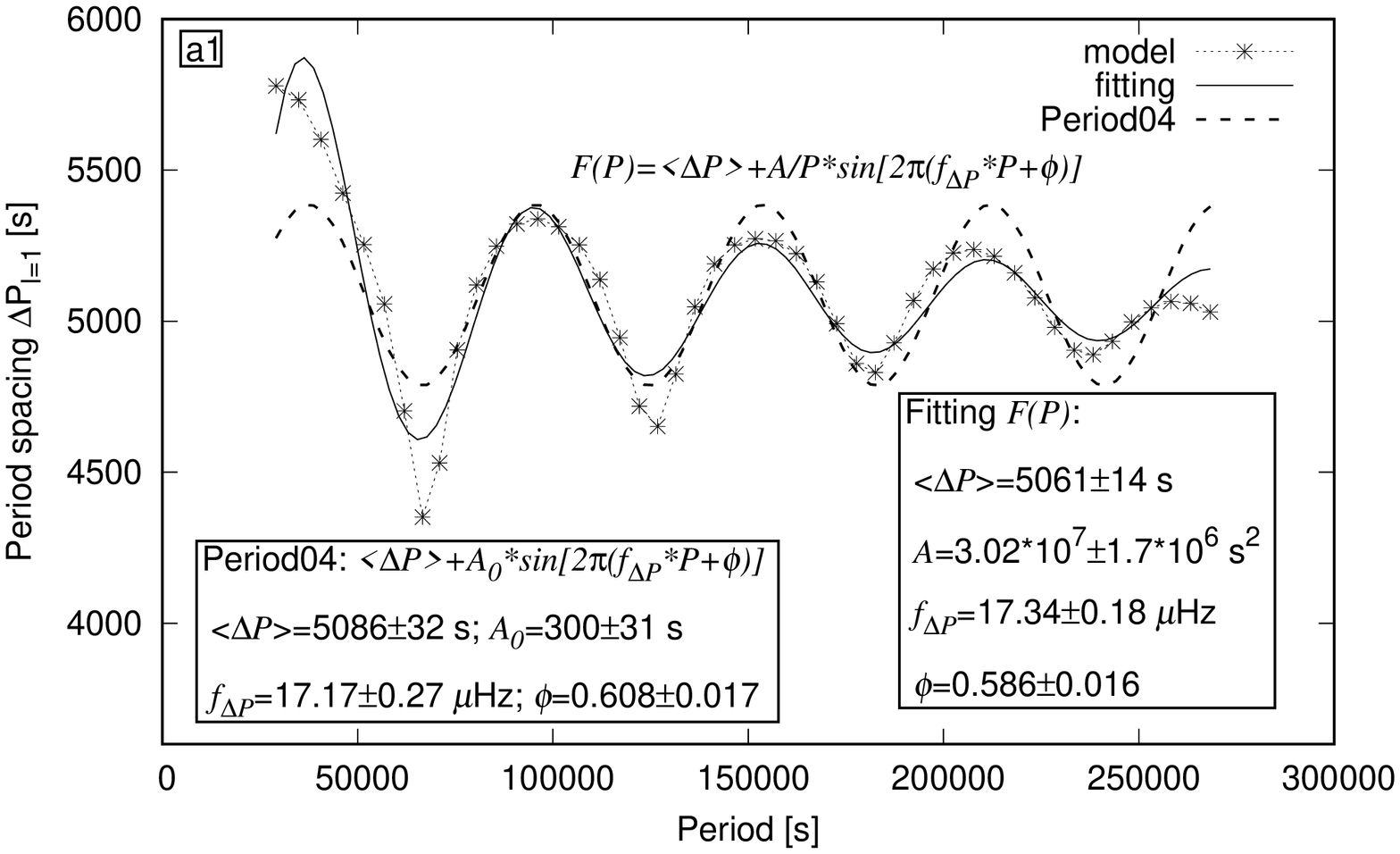}
\includegraphics[scale=0.45,angle=-90]{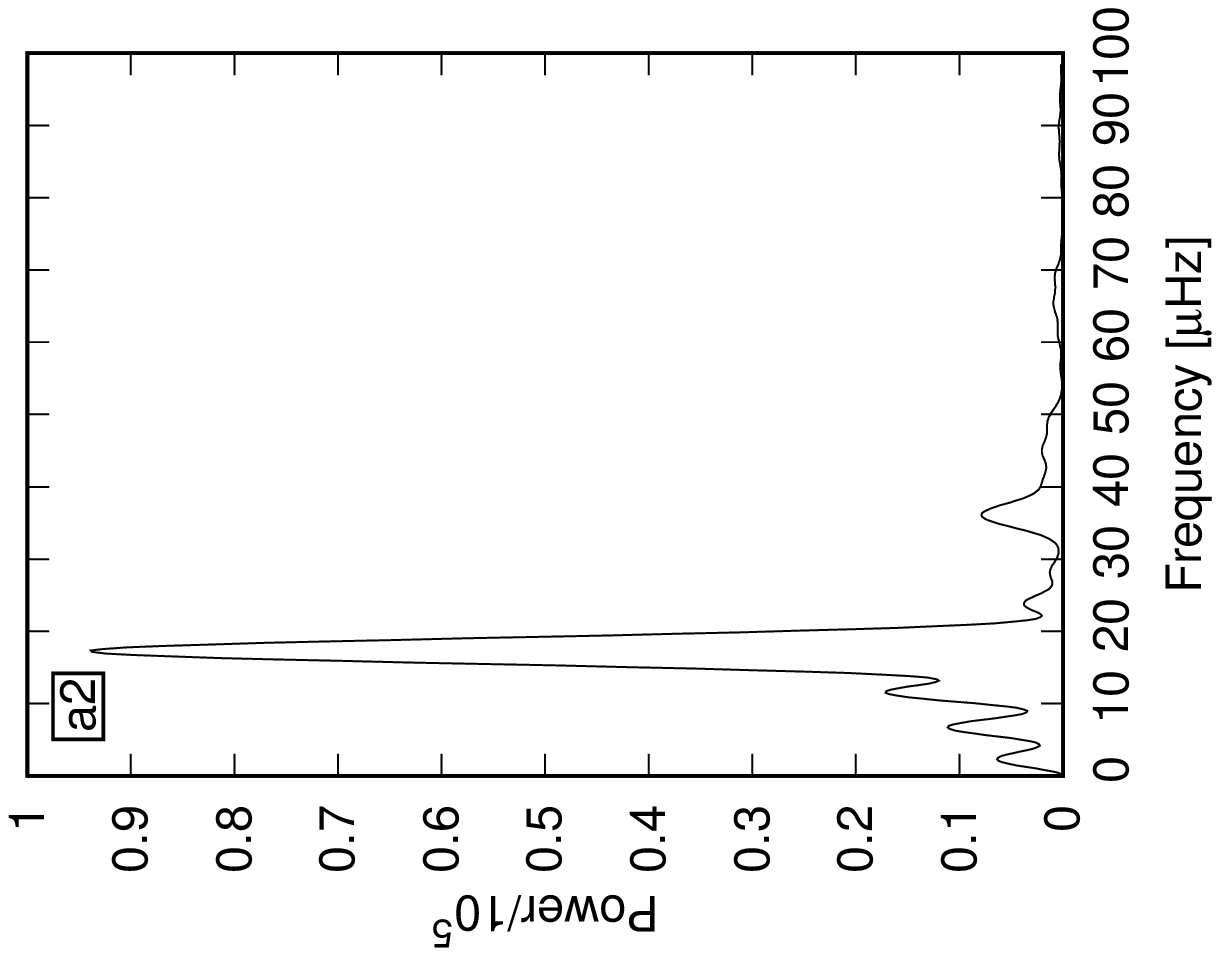}\\
\includegraphics[scale=0.45,angle=-90]{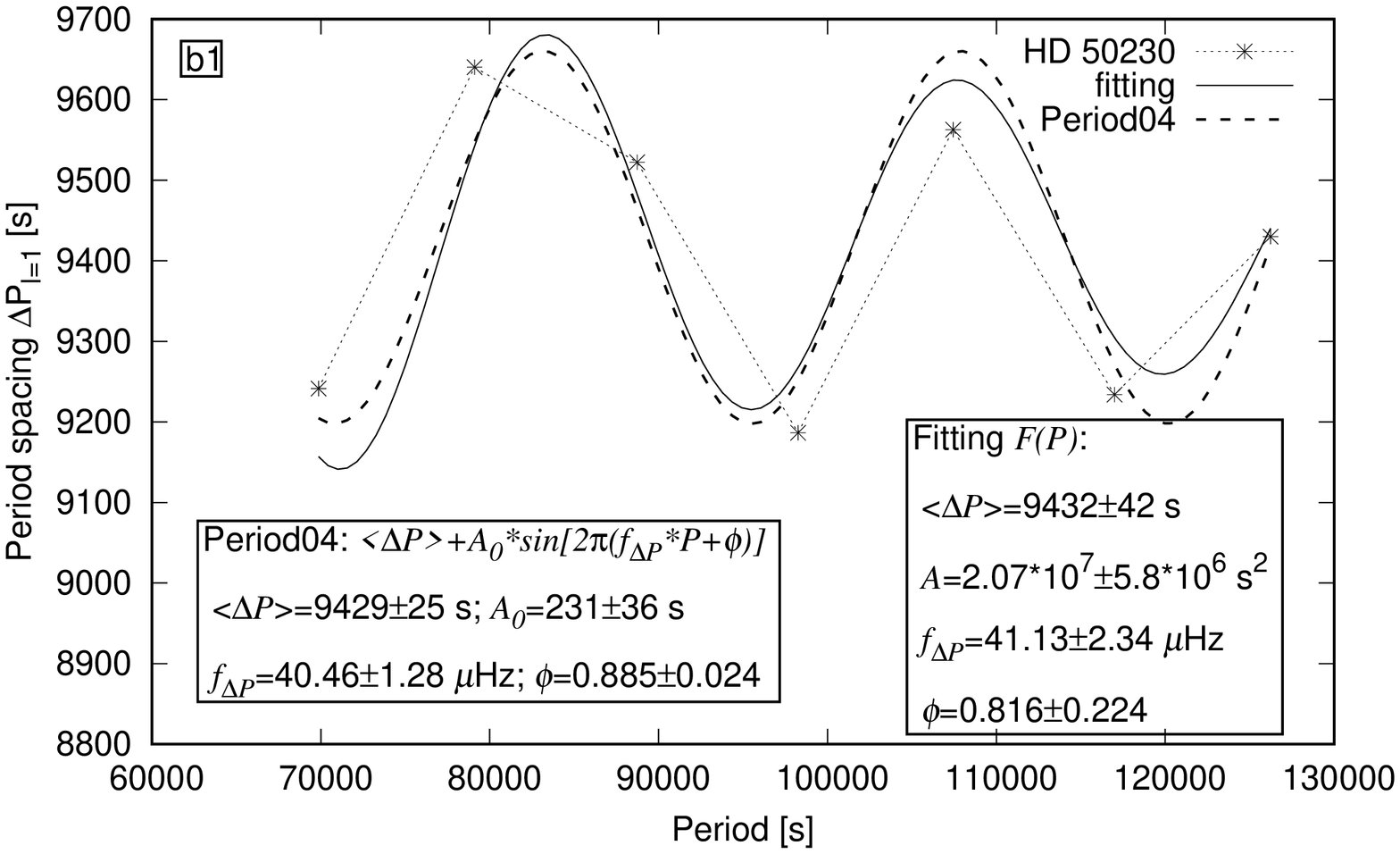}
\includegraphics[scale=0.45,angle=-90]{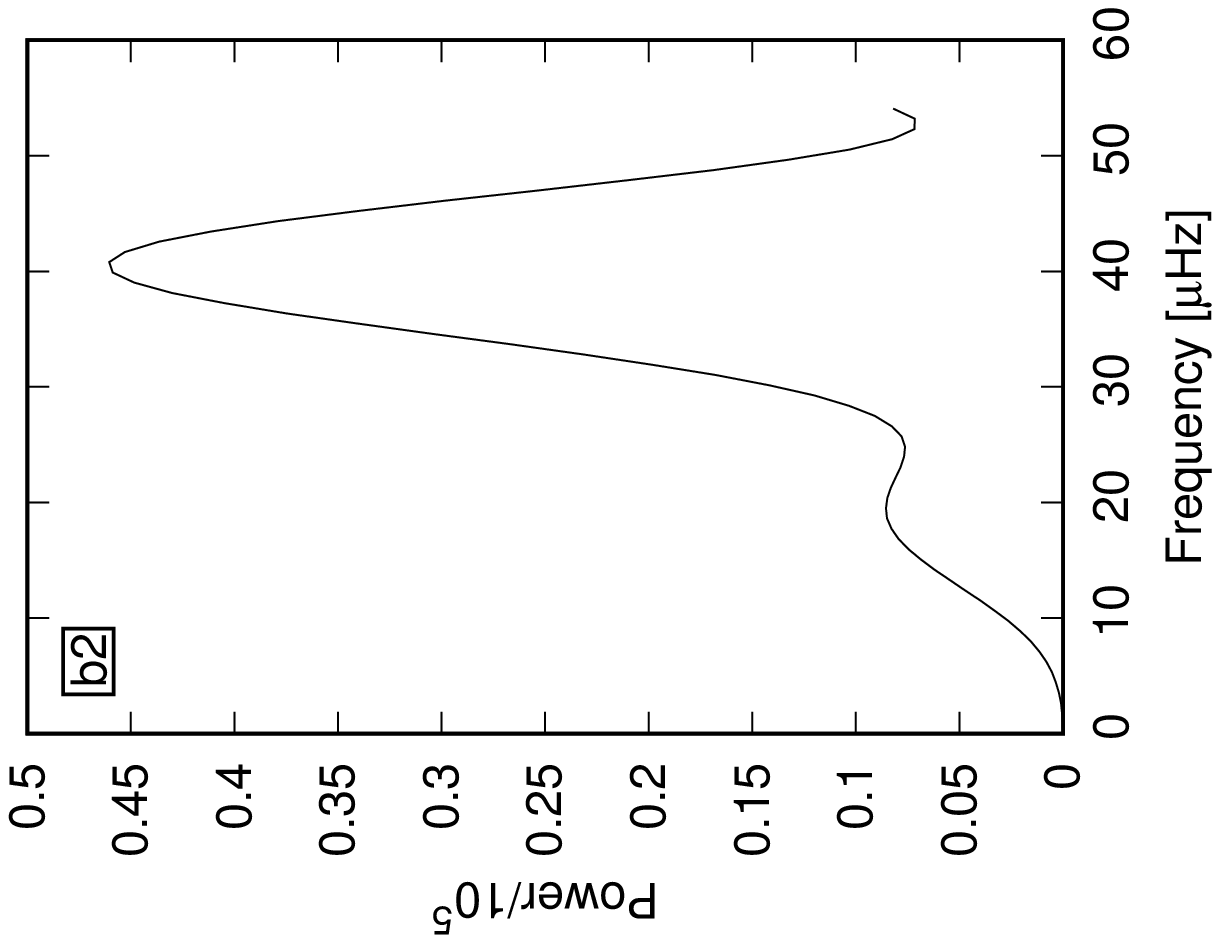}
   \caption{The analyzing results of period spacings against periods for model calculations (with the fundamental parameters of $M=3.0~{M_{\odot}}$, $Z_{\rm init}=0.007$, $f_{\rm ov}=0.01$, $\log D_{\rm mix}=1.0$, and $X_{\rm C}\simeq0.60$; Upper Panels) and the observations (HD 50230; Bottom Panels), respectively. The right-hands panels correspond to the fourier transformations of the left panels with software Period04. In left-hands figures, solid- and dashed-lines represent the results via fitting with Equation \eqref{eq:fit} and the analysis of software Period04, respectively. The corresponding parameters are marked in figures. Correspondingly, the analyzing results of the other three observations (KIC 6462030, KIC 8324482, KIC 10526294, and KIC 10526294$^{\rm a}$ (the largest five periods of KIC 10526294 are removed from the period series)) are shown in Figure \ref{fig.A1}. Parts of the analyzing parameters ($f_{\Delta P}$ and $\langle\Delta P\rangle$) are listed in Table \ref{table_1}.
  }\label{fig.fit1}
  \end{center}
\end{figure*}

\begin{figure}
  \begin{center}
\includegraphics[scale=0.43,angle=-90]{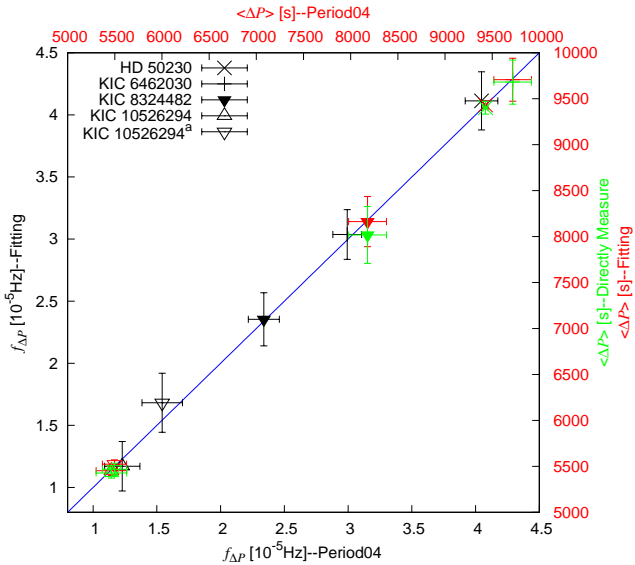}
   \caption{ Comparisons between different methods for $f_{\Delta P}$ and $\langle\Delta P\rangle$ of observations. They are listed in Table \ref{table_1}. Different point types present different targets and different colors correspond different parameters. Black points --- fitted $f_{\Delta P}$ vs. Period04 analyzed $f_{\Delta P}$; red points --- fitted $\langle\Delta P\rangle$ vs. Period04 analyzed $\langle\Delta P\rangle$; green points --- directly estimated $\langle\Delta P\rangle$ vs. Period04 analyzed $\langle\Delta P\rangle$. KIC 10526294$^{\rm a}$ denotes that the largest five periods are removed from the period series of KIC 10526294. It will be used in the following text.
  }\label{fig.obs}
  \end{center}
\end{figure}

\begin{deluxetable*}{lccccccccc}
\centering
\tablecaption{Summary of the Analysis for Observations, including mean period spacing $\langle\Delta P\rangle$, variational frequency of period spacing $f_{\Delta P}$, and their rotational parameters (rotational period $P_{\rm rot}$, rotational frequency $f_{\rm rot}$, or velocity $v\sin i$). \label{table_1}}
\tablehead{
  & \multicolumn{2}{c}{Period04 extracted} & \multicolumn{2}{c}{Fitting with Equation \eqref{eq:fit}}  \\
\colhead{ID} &  \colhead{$f_{\Delta P}$} & \colhead{$\langle\Delta P\rangle$} & \colhead{$f_{\Delta P}$} & \colhead{$\langle\Delta P\rangle$} &  \colhead{$\langle\Delta P\rangle^{\rm a}$}  & \colhead{$P_{\rm rot}^{\rm b}$, $f_{\rm rot}^{\rm c}$}   &\colhead{Obs. \& Rot.}   & \colhead{Symb.} & \colhead{$\Lambda_{\mu}$} \\
\colhead{} & \colhead{[$10^{-5}$Hz]} &\colhead{[s]} & \colhead{[$10^{-5}$Hz]} &\colhead{[s]}  & \colhead{[s]}& \colhead{ or $v\sin i^{\rm d}$} &\colhead{Ref.} & \colhead{in Fig.} & \colhead{[$10^{-5}$Hz]}
}
\startdata

HD 50230     &4.046$\pm$0.128  &9429$\pm$25  & 4.113$\pm$0.234  &  9432$\pm$42  &  9402$\pm$69 & 0.044$\pm$0.007$^{\rm c}$ & 1 & {$\mathbf \times$}  & 8.08$\pm$0.07 \\
     &   &    &   &  &  &  6.9$\pm$1.5$^{\rm d}$ & 1 &   &  \\

KIC ~6462030 &2.993$\pm$0.113  &  9717$\pm$200 &  3.037$\pm$0.200  &  9707$\pm$234  & 9681$\pm$240 & 4.6$\pm$3.0$^{\rm d}$ & 2 & {\large\bf +} &  \\
KIC ~8324482 & 2.339$\pm$0.122  &  8179$\pm$201  &  2.354$\pm$0.214  &  8163$\pm$273    & 8018$\pm$310 & $-$18.5$\pm$2.4$^{\rm d}$ & 2 & $\blacktriangledown$  & 5.61$\pm$0.09       \\
KIC 10526294$^{\rm e}$ & 1.229$\pm$0.138  &  5464$\pm$162    & 1.171$\pm$0.199$^{\rm g}$  &  5453$\pm$52$^{\rm g}$  & 5428$\pm$56 & $\sim$188$^{\rm b}$& 3 & $\bigtriangleup$ & 3.51$\pm$0.07 \\
KIC 10526294$^{\rm f}$ & 1.542$\pm$0.159  &  5497$\pm$128    & 1.682$\pm$0.238$^{\rm g}$  &  5523$\pm$48$^{\rm g}$  & 5469$\pm$66 & & & $\bigtriangledown$ & 3.51$\pm$0.07 
\enddata
\tablenotetext{}{Obs. \& Rot. Ref. (The references of the observational data and the rotational parameters): 1--\citet[][]{Degroote2012AA}; 2--\citet[][]{Zhang2018ApJ}; 3--\citet[][]{Papics2014AA}.
}
\tablenotetext{a}{Directly calculating the mean period spacing $\langle\Delta P\rangle$ and using the standard error as its measuring uncertainty.}
\tablenotetext{b}{Rotational period with the unit of day.}
\tablenotetext{c}{Rotational frequency with the unit of ${\rm d}^{-1}$.}
\tablenotetext{d}{Rotational velocity with the unit of km s$^{-1}$.}
\tablenotetext{e}{Analyzing with all of 19 observed periods, i.e., all of 18 period spacings.}
\tablenotetext{f}{Analyzing with 14 observed periods, i.e., 13 period spacings. The largest five periods are removed from the period series. In the work, we denote it as KIC 10526294$^{\rm a}$.}
\tablenotetext{g}{The errors are determined with the method of Least-Squares Fitting. The others are determined with 1000 times Monte Carlo Simulations.}
\end{deluxetable*}

The asteroseismic analysis suggested that the convective overshooting scale out of the convective core are mainly in range of 0.1 -- 0.3 $H_{P}$ (local pressure scale heights) for $\beta$ Cephei stars \citep[$M\gtrsim8~\rm M_{\odot}$; see discussions in][(Table 1 for a review) for exapmle]{Saio2012ASPC}. \citet[][]{Degroote2010Natur} analyzed the period spacings of SPB star HD 50230 ($M=7-8~\rm M_{\odot}$) and suggested that the overshooting extent of the convective core is about 0.2 -- 0.3 $H_{P}$. \citet[][]{Papics2014AA} analyzed the period spacings of KIC 10526294 ($M=3.25~\rm M_{\odot}$) and suggested that the core overshooting is less than or equal to $0.15~H_{P}$ with step function overshooting and $f_{\rm ov}\lesssim0.015$ with exponentially decreasing overshooting. Hereafter, \citet[][]{Moravveji2015AA} also analyzed KIC 10526294 with step function and exponentially decreasing overshooting and reported that the exponentially decreasing overshooting is better than the step function overshooting for interpreting the observations and obtained the overshooting parameter $f_{\rm ov}=0.017-0.018$. \citet[][]{Moravveji2016ApJ} analyzed the faster rotator KIC 7760680 and claimed that the overshooting parameter $f_{\rm ov}\approx0.024\pm0.001$. In the work of \citet[][]{Deng2018}, the best value of the overshooting parameter is $f_{\rm ov}=0.03$ in KIC 8324482. In addition, the optimal overshooting parameter of HD 50230 is about $f_{\rm ov}=0.0175-0.020$ \citep[][]{Wu2018}. Based on these previous works, we set the overshooting parameter $f_{\rm ov}\in[0.01,~0.03]$ with a step of 0.01 in our theoretical models.

In order to perfectly interpret the observations (the period spacings and their variations), the extra diffusion mixing ($D_{\rm mix}$ or $D_{\rm ext}$) should be considered in the theoretical models \citep[e.g.,][]{Degroote2010Natur,Moravveji2015AA,Moravveji2016ApJ}. It mainly works in the radiative zone above the convective core to slightly smooth the $\mu$-gradient at the region of $\nabla_{\mu}$ rapidly decreasing and closing to zero.

In the work of \citet[][]{Degroote2010Natur}, \citet[][]{Moravveji2015AA}, and \citet[][]{Moravveji2016ApJ}, they suggested that the best values of the extra diffusion mixing parameter $\log D_{\rm mix}$ are 3.4 -- 4.3 in HD 50230, 1.75 -- 2.00 in KIC 10526294, and $0.75\pm0.25$ in KIC 7760680, respectively. In the present work, we set the extra diffusion mixing parameter $\log D_{\rm mix}\in[1.0,~3.0]$ with a step of 1.0 for all of the calculated masses.

Similar to the works of \citet[][]{Moravveji2015AA} and \citet[][]{Moravveji2016ApJ}, we set the initial hydrogen mass fraction $X_{\rm init}=0.71$ taken from the Galactic B-star standard \citep[][]{Nieva2012AA}. The initial metal mass fractions $Z_{\rm init}$ are set as 0.007 (poor metal), 0.014 (near solar), and 0.028 (richer metal), respectively. Surely, the initial helium mass fractions are $Y_{\rm init}=1-X_{\rm init}-Z_{\rm init}$. The initial mass $M_{\rm init}$ ranges from 3.0 to 8.0 $\rm M_{\odot}$ with a step of 1.0 $\rm M_{\odot}$.


\section{Results}\label{sec.result}

According to the theoretical deduction (Equations \eqref{eq:sindp} and \eqref{eq:Ap}), model calculations, and observations, we adopt the following sinusoidal function to fit the period spacings against periods for both of observations and model calculations and further to decide the variational frequency of period spacing $f_{\Delta P}$ and the mean period spacings $\langle\Delta P\rangle$, i.e.,
\begin{equation}\label{eq:fit}
\Delta P_{n}=\langle\Delta P\rangle+\frac{A}{P_{n}}\cos[2\pi(f_{\Delta P}*P_{n}+\phi)].
\end{equation}

As a matter of fact, as shown in Figure \ref{fig.1} the variation of period spacings becomes more and more frequently and the fitting will become more and more difficult when the center hydrogen mass fraction $X_{\rm C}$ becomes lower and lower. In order to conveniently decide variational frequency $f_{\Delta P}$ and mean period spacing $\langle\Delta P\rangle$, we can directly make Fourier transformation for the period spacings with the software Period04 \citep[e.g.,][]{Lenz2004IAUS..224..786L,Lenz2005CoAst.146...53L,Lenz2014ascl.soft07009L}, i.e., fitting with the relation of $\Delta P_{n}=\langle\Delta P\rangle+A_{0}\cos[2\pi(f_{\Delta P}*P_{n}+\phi)]$. On the other hands, we can also directly calculate the mean period spacing $\langle\Delta P\rangle$ from all of the calculated or observed period spacings $\Delta P_{n}$ and use its standard errors as the measure uncertainty.

Essentially, the three different methods are similar but with different formulas. For the former two ways, they use sinusoidal function to fit period spacings, but their amplitude are variable in Equation \eqref{eq:fit} and a constant in Period04 analyzing, respectively. Compared to the former two ways, the least one just estimate the zero point and the periodical signal is ignored. In other words, the formula of these relations gradually degenerate and simplify among the three methods.

The fitting results are shown in Figures \ref{fig.fit1}, \ref{fig.A1}, and \ref{fig.obs} and listed in Table \ref{table_1} for the observations. In Figures \ref{fig.fit1} and \ref{fig.A1}, the fitting parameters and the Period04 analyzed coefficients are symbolled with text. The corresponding coefficients are consistent between them. In addition, the fitting result with Equation \eqref{eq:fit} against Period04 analyzed are shown in Figure \ref{fig.obs} with black points for $f_{\Delta P}$ and red points for $\langle\Delta P\rangle$, respectively. In addition, the directly estimated $\langle\Delta P\rangle$ against Period04 analysed $\langle\Delta P\rangle$ is also shown in Figure \ref{fig.obs} with green points. It can be found from these figures and Table \ref{table_1} that the results are well consistent with each other. In the present work, we adopt Period04 software to extract the frequency $f_{\Delta P}$ and directly calculate the average value of the period spacings $\langle\Delta P\rangle$. In addition, the measure uncertainty of $f_{\Delta P}$ is determined with 1000 times Monte Carlo Simulation in software Period04 and that of mean period spacings $\langle\Delta P\rangle$ is characterized with the standard errors of $\langle\Delta P\rangle$.

As shown in Figures 4, A.1, and A.2 of \citet[][]{Moravveji2015AA} the observations of KIC 10526294 are well consistent with the best fitting model for the low-period modes ($P\lesssim1.75$ day). For the longer period modes, the tendencies of period spacings are different between the observations and models. In the present work, we analyze them with two different ways. The first, we analyze all of those observed modes. The second, we remove the five largest period modes and leave 13 modes and analyze them. It is noted with KIC 10526294$^{\rm a}$. Their fitting results are shown in Figures \ref{fig.A1} and listed in Table \ref{table_1}. It can be seen from the fitting results that the latter has larger $f_{\Delta P}$. The corresponding fitting curve is closer to the theoretical model in the tendency of period spacings compared the fitting results (Figure \ref{fig.A1}) with modelling \citep[Figures 4, A.1, and A.2 of][]{Moravveji2015AA}. In addition, as shown in Figures \ref{fig.4} and \ref{fig.5} the latter will predict an older star, i.e., smaller hydrogen in center.

Notes that except KIC 10526294 we also use the three stars -- HD 50230, KIC 6462030 and KIC 8324482 -- as examples in the present study. They are non-rotator and/or slow-rotator. In addition, they present periodic variation in the period spacing series as shown in theoretical predicted (see Figure \ref{fig.1}). For KIC 8324482 and KIC 6462933, \citet[][]{Zhang2018ApJ} extracted individual periods and determined their period spacings. Those modes seem to be $l=1$ and $m=0$ (private communicate with Zhang). \citet[][]{Zhang2018ApJ} does not give clearly mode identification in their paper, since there is not seismic modelling in details. For KIC 8324482, \citet[][]{Deng2018} made seismic analysis and found that those observed modes can be fitted with $l=1$ and $m=0$ modes. In the work of \citet[][]{Wu2018}, they seismically modelled HD 50230 in details and found that the 8 observed modes by \citet[][]{Degroote2010Natur,Degroote2012AA} are $l=1$ and $m=0$. For KIC 6462933, the modelling are on going, except two period spacings have larger discrepancy between observations and models the other modes seems to be fitted with $l=1$ and $m=0$ modes. Therefore, in the present work, we analyze the four stars --HD 50230, KIC 6462030, KIC 8324482, and KIC 10526294 -- with $l=1$ and $m=0$.

\begin{figure}
  \begin{center}
\includegraphics[scale=0.43,angle=-90]{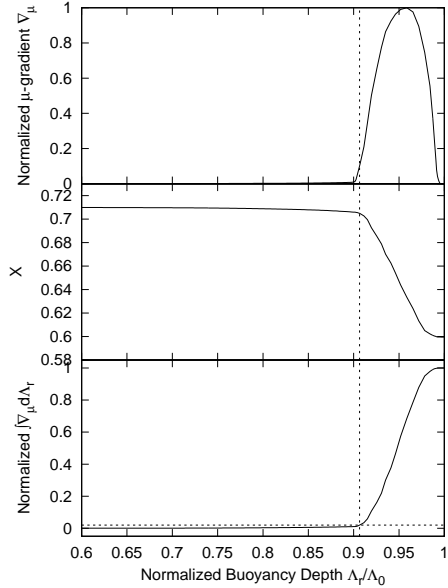}
   \caption{The profiles of normalized $\mu$-gradient $\nabla_{\mu}$ (upper panel), hydrogen mass fraction $X$ (middle panel), and of normalized integral $\int^{\Lambda_{r}}_{0}\nabla_{\mu}d\Lambda_{r'}$ (bottom panel), respectively. The corresponding period spacings are shown in Panel (a1) of Figure \ref{fig.fit1}. Its fundamental parameters are $M=3.0~{M_{\odot}}$, $Z_{\rm init}=0.007$, $f_{\rm ov}=0.01$, $\log D_{\rm mix}=1.0$, and $X_{\rm C}\simeq0.60$. In panels, the vertical dashed-line corresponds the outer boundary of $\mu$-gradient region, i.e., $r_{\mu}$. The horizontal dashed-line in bottom panel presents $y=0.02$.
  }\label{fig.pro}
  \end{center}
\end{figure}

As the definition of $\Lambda_{\mu}$ in Equation \eqref{eq:Lambdamu}, the value of $\Lambda_{\mu}$ is decided by the outer boundary of $\mu$-gradient region ($r_{\mu}$) and the corresponding $N$. As shown in Figure \ref{fig.1} the $r_{\mu}$ is defined at the beginning of the hydrogen decreasing, i.e., $\nabla_{\mu}>0$. If the outer envelope is fully homogeneous composition, $r_{\mu}$ can be clearly decided at the position of $\nabla_{\mu}\neq0$ from stellar surface to center. However, for real stars, especially for those stars which have an extra-mixing diffusion on radiative zone, $\Lambda_{\mu}$ correspondingly has a small value in the whole radiative zone beyond the inner $\mu$-gradient region as shown in Figure \ref{fig.pro}. And then deciding the outer boundary of $\mu$-gradient region ($r_{\mu}$) becomes more and more difficult.

According to the theory of stellar structure and evolution, the profile of $\nabla_{\mu}$ beyond convective core varies with stellar evolution, stellar mass, overshooting extension, and extra-mixing diffusion. Therefore, giving a certainty value for $\nabla_{\mu}$ and cutting $r_{\mu}$ to estimate $\Lambda_{r}$ is unsuitable. In order to conveniently estimate $\Lambda_{\mu}$ from stellar models, we introduce an integral -- $\int^{\Lambda_{r}}_{0}\nabla_{\mu}d\Lambda_{r'}$ -- as a criterion. We assume 98 percents of the integral to be within the decided $\mu$-gradient region as shown in the bottom panel of Figure \ref{fig.pro}, i.e., the contribution of integral is less than 2 percents in the whole similar homogeneous composition radiative envelope. For most of the calculated models, the contributions of the outer region for the integral are around the level. Therefore, we adopt the normalized integral $\int^{\Lambda_{0}}_{\Lambda_{r}}\nabla_{\mu}d\Lambda_{r'}/\int^{\Lambda_{0}}_{0}\nabla_{\mu}d\Lambda_{r'}=0.02$ as the outer boundary of $\mu$-gradient region ($r_{\mu}$) to estimate $\Lambda_{\mu}$.

Correspondingly, we adopt the different of $\Lambda_{r}$ between adjacent points at the position of $r_{\mu}$ as the measure uncertainty for $\Lambda_{\mu}$. Such measure uncertainty just represents the density of the mesh grid for stellar models. They are on the level of $0.5-2\times10^{-6}$ Hz. As shown in Figure \ref{eq:fit} the error-bars are less than the size of data points.

For such method, $\Lambda_{\mu}$ and their measure uncertainties are conveniently determined from theoretical models. But, it will be slightly partly underestimated $\Lambda_{\mu}$ for some cases, such as when the star has lower mass but with a larger age and larger extra diffusion mixing parameter ($\log D_{\rm mix}$).

\subsection{$f_{\Delta P}$ vs. $\Lambda_{\mu}$}

\begin{figure}
  \begin{center}
\includegraphics[scale=0.35,angle=-90]{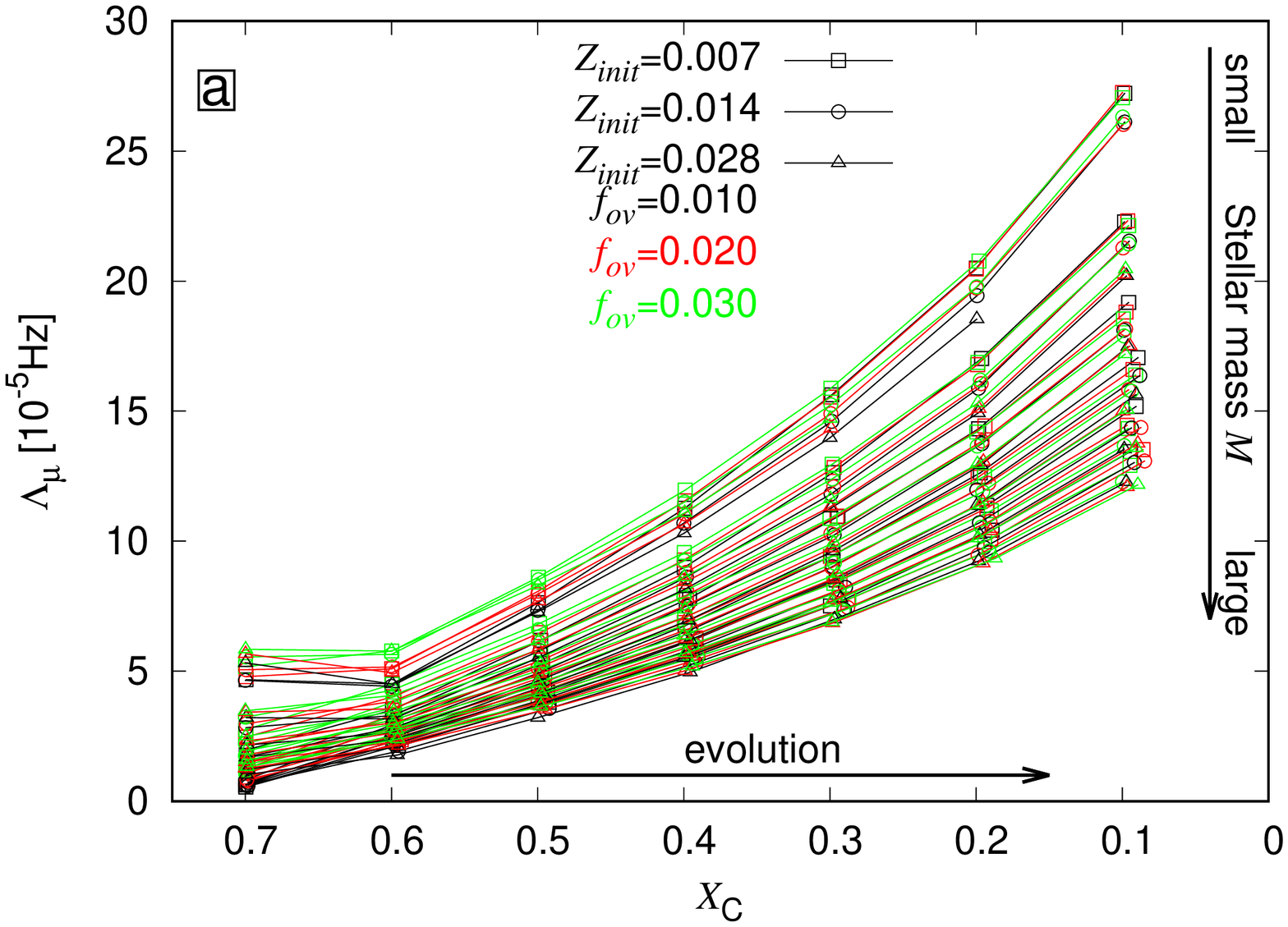}
\includegraphics[scale=0.35,angle=-90]{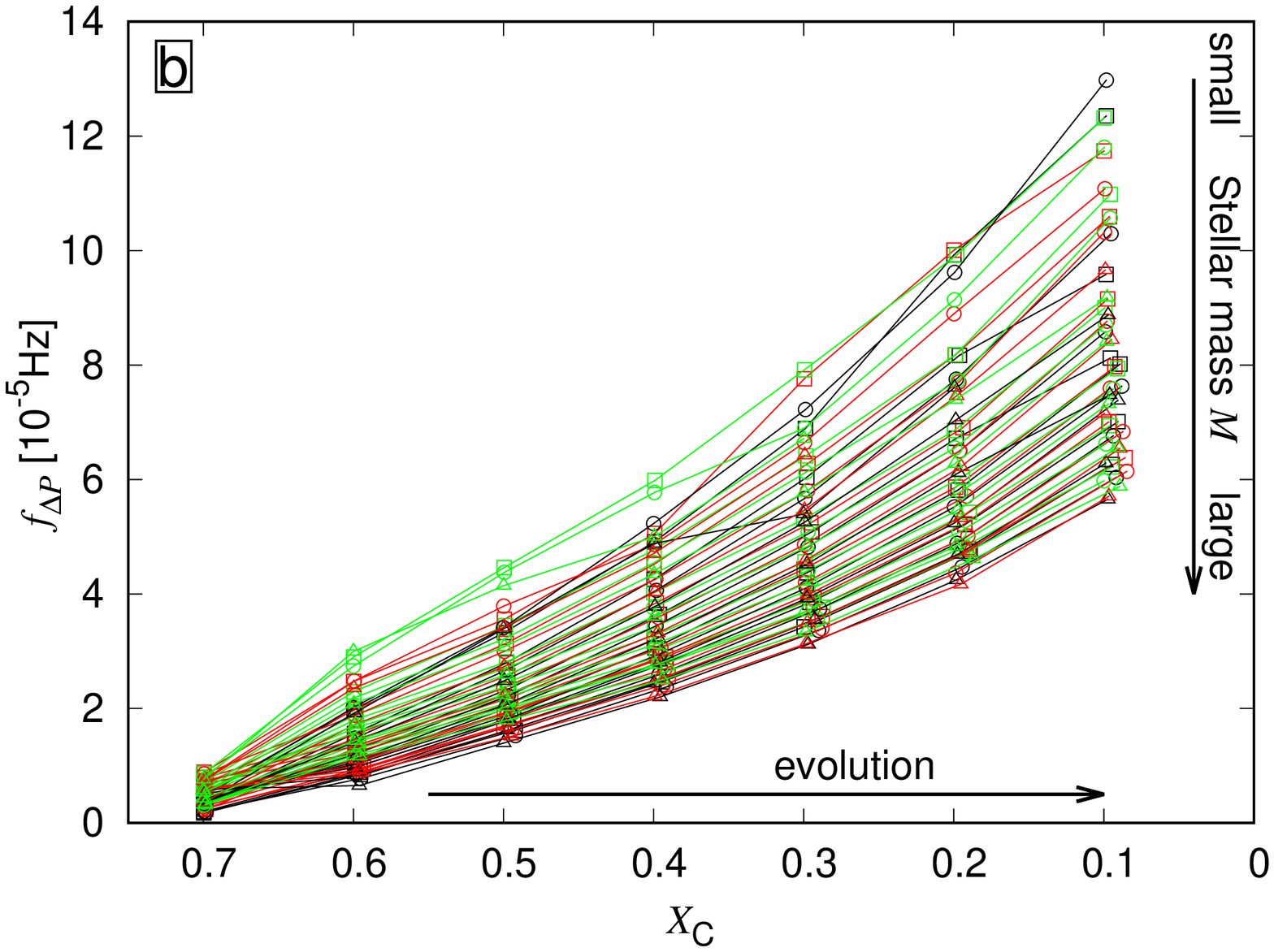}
\includegraphics[scale=0.35,angle=-90]{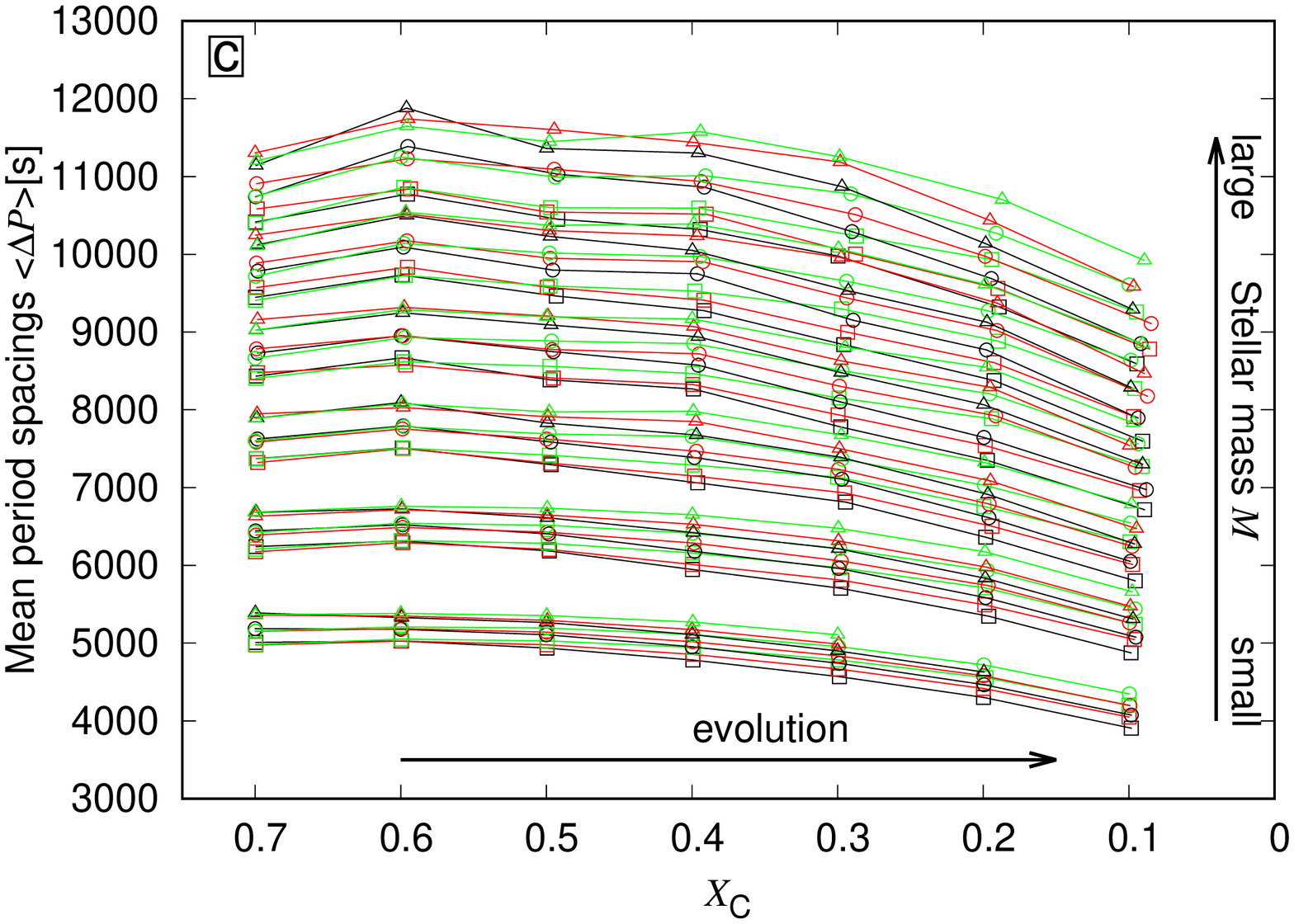}
\caption{{\bf (a):} Buoyancy size ($\Lambda_{\mu}$) as a function of the central mass fraction of hydrogen ($X_{\rm C}$); {\bf (b):} Similar to panel (a), but for the frequency of the variations of period spacings ($f_{\Delta P}$); {\bf (c):} $\langle\Delta P\rangle$ vs. $X_{\rm C}$. In panels (a) -- (c), all of the models with the same extra diffusion coefficient ($\log D_{\rm mix}=2.0$), different initial metal mass fraction ($Z_{\rm init}$), and different overshooting extension ($f_{\rm ov}$). They are represented with different point types and colors: open circles --- $Z_{\rm init}=0.007$; open squares --- $Z_{\rm init}=0.014$; open upper triangles --- $Z_{\rm init}=0.028$; black --- $f_{\rm ov}=0.010$; red --- $f_{\rm ov}=0.020$; and green --- $f_{\rm ov}=0.030$, respectively. In addition, the horizontal arrows point the direction of evolution. The vertical arrows denote the stellar mass ($M$). It varies from small ($3.0~{\rm M_{\odot}}$) to large ($8.0~{\rm M_{\odot}}$) along with the direction of arrows.
  }\label{fig.2}
  \end{center}
\end{figure}

\begin{figure}
  \begin{center}
\includegraphics[scale=0.35,angle=-90]{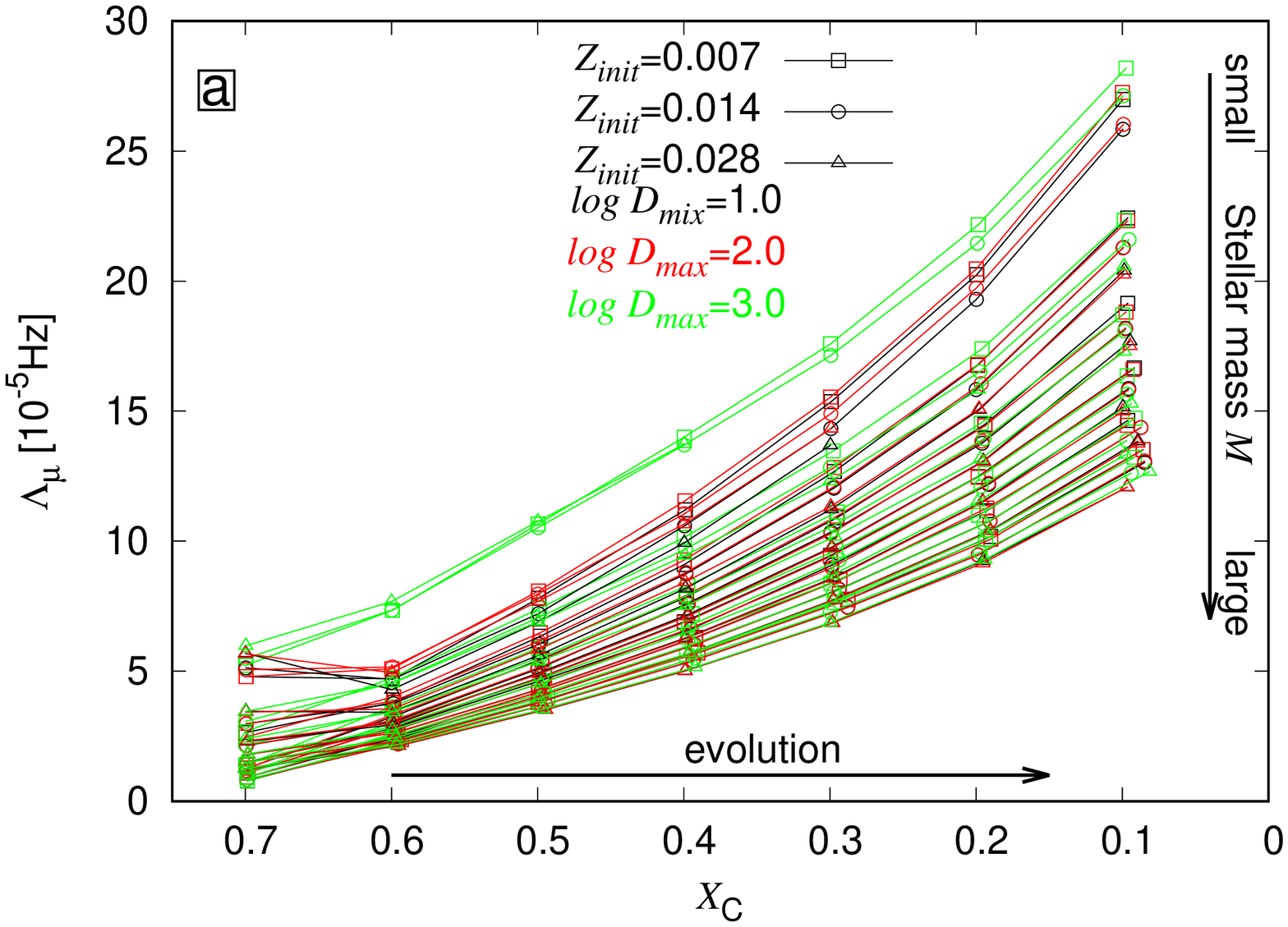}
\includegraphics[scale=0.35,angle=-90]{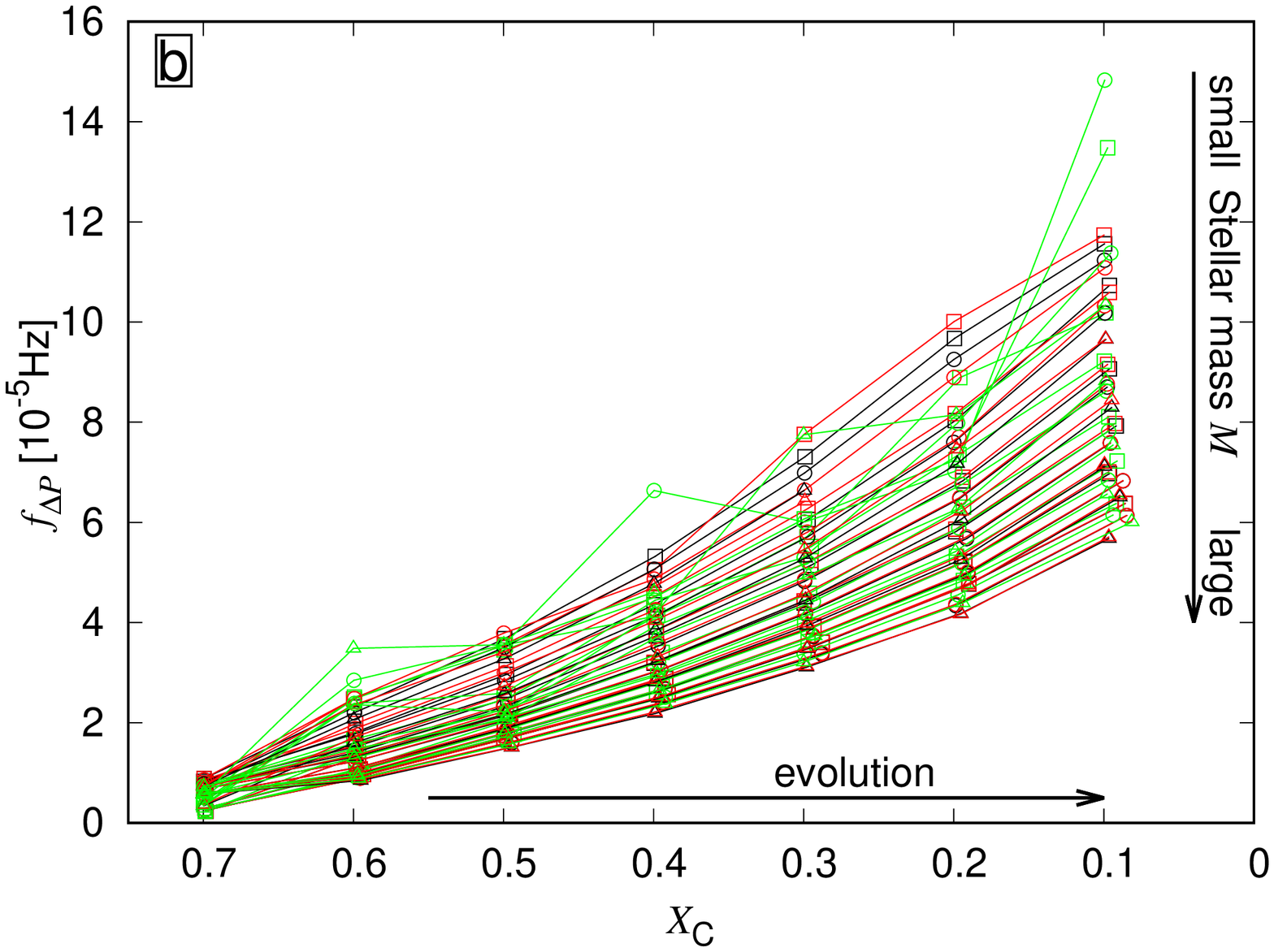}
\includegraphics[scale=0.35,angle=-90]{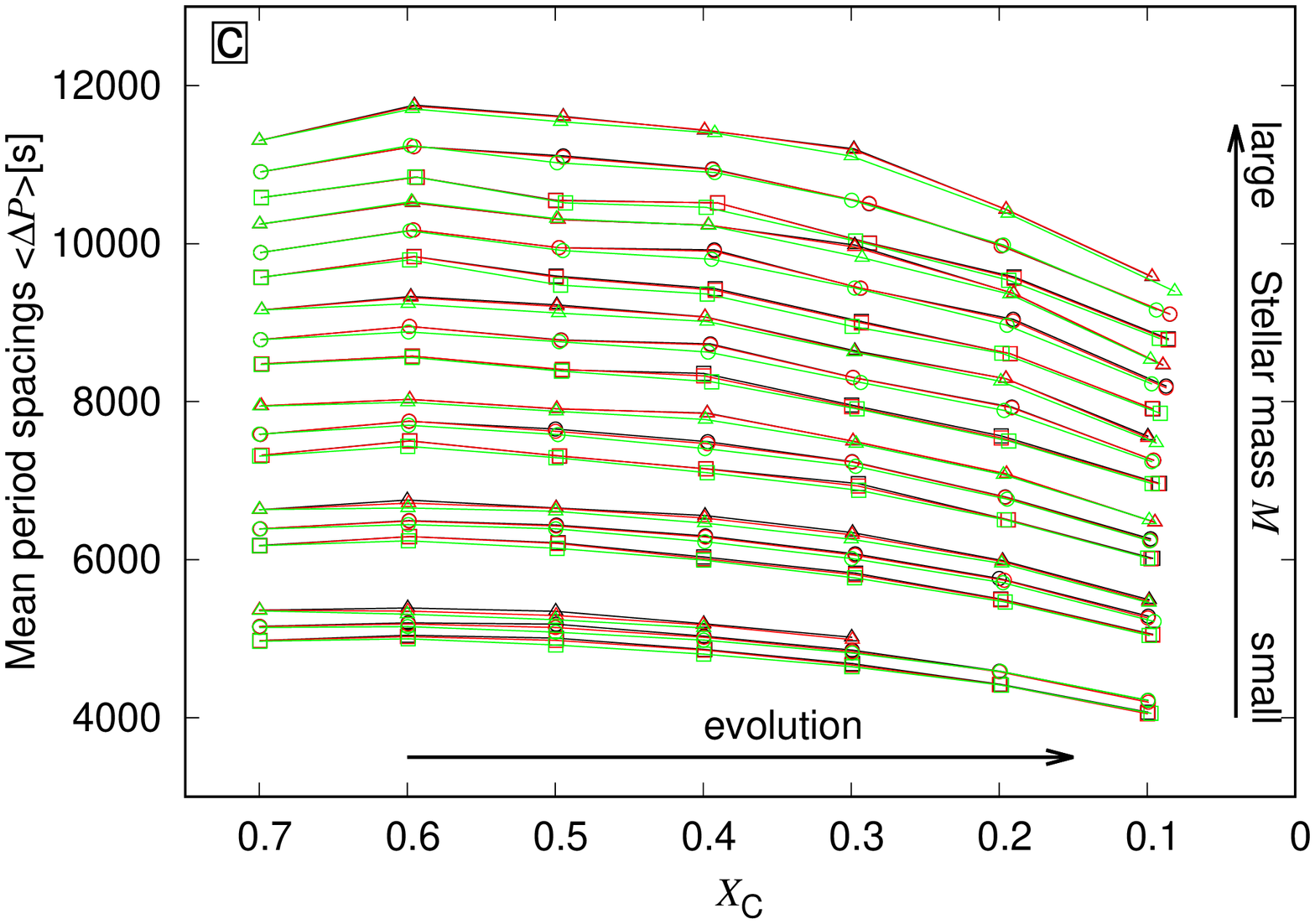}
\caption{Similar to Figure \ref{fig.2}, but for being fixed overshooting extend in the convective core and different extra-mixing parameters with different colors similarly. They are $f_{\rm ov}=0.020$ and $\log D_{\rm mix}=1.0$ (black), 2.0 (red), and 3.0 (green), respectively.
  }\label{fig.2.1}
  \end{center}
\end{figure}

\begin{figure}
  \begin{center}
\includegraphics[scale=0.43,angle=-90]{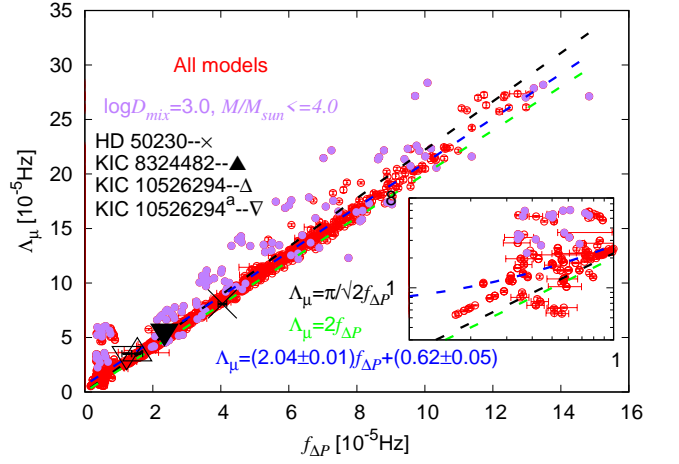}
\caption{$\Lambda_{\mu}$ vs. $f_{\Delta P}$ for all of the calculated models with $M\in[3,~8]$ with a step of 1.0 $\rm M_{\odot}$, $Z_{\rm init}\in[0.007,~0.014,~0.028]$, $f_{\rm ov}\in[0.010,~0.020,~0.030]$, and $\log D_{\rm mix}\in[1.0,~2.0,~3.0]$. Those points are shown with error-bar. For most of models, the error-bar are smaller than the size of the points. The filled purple circles represent the models whose extra diffusion coefficient $\log D_{\rm mix}$ equals 3.0 and mass $M/{\rm M_{\odot}}<=4$. The four black points denote the ``Observations": HD 50230, KIC 8324482, KIC 10526294, and KIC 10526294$^a$, respectively. For the three dashed-lines: black presents the theoretical prediction --- Equation \eqref{eq:Lfp} --- $\Lambda_{\mu}=\frac{\pi}{\sqrt{2}}f_{\Delta P}\simeq2.22f_{\Delta P}$; green --- $\Lambda_{\mu}=2f_{\Delta P}$; and blue --- the fitting.
  }\label{fig.2.2}
  \end{center}
\end{figure}

The calculations and fitting results are shown in Figures \ref{fig.2}-\ref{fig.2.2}. Figures \ref{fig.2}(a) and \ref{fig.2.1}(a) illustrate the relationship between the width of $\mu$-gradient region in buoyancy radius $\Lambda_{\mu}$ and the central hydrogen mass fraction $X_{\rm C}$. The variational frequency of the period spacing $f_{\Delta P}$ and the mean period spacings $\langle\Delta P\rangle$ varying with $X_{\rm C}$ are shown in Figures \ref{fig.2}(b) and \ref{fig.2.1}(b) and Figures \ref{fig.2}(c) and \ref{fig.2.1}(c) respectively. In addition, $f_{\Delta P}$ vs. $\Lambda_{\mu}$ is shown in Figure \ref{fig.2.2}. 

As shown in Figures \ref{fig.2}(a)-(b) and \ref{fig.2.1}(a)-(b), $f_{\Delta P}$ and $\Lambda_{\mu}$ increase with the decreasing $X_{\rm C}$. It indicates that they are sensitive to the central status of SPB stars \citep[also refer to e.g.,][]{Miglio2008MNRAS,Moravveji2015AA}. On the other hand, it can be seen from Figures \ref{fig.2}(a)-(b) and \ref{fig.2.1}(a)-(b) that $f_{\Delta P}$ and $\Lambda_{\mu}$ decrease with the increasing of stellar mass ($M$) for a given value of the central hydrogen mass fraction $X_{\rm C}$. In addition, the period spacing $\Delta P$ varies with central hydrogen mass fraction $X_{\rm C}$ and increases with increasing $M$ \citep[also refer to e.g.,][]{Miglio2008MNRAS,Moravveji2015AA,Ouazzani2017MNRAS,Ouazzani2018arXiv}.

Similar to Equation \eqref{eq:fit}, the theoretically expected relation (i.e., Equation \eqref{eq:sindp}) can be expressed as the form of:
\begin{equation}\label{eq:sindp1}
\delta P_{n}\propto A_{\delta P}\cos[2\pi (\frac{L\Lambda_{\mu}}{\pi}\ast P_{n}+\phi')],
\end{equation}
where the phase $\phi'=1/4$ compared with Equation \eqref{eq:sindp}.

Compared theoretically expected relation (Equation \eqref{eq:sindp1}) with the fitting relation (Equation \eqref{eq:fit}), we can easily obtain that
\begin{equation}\label{eq:Lfp}
\Lambda_{\mu}=\frac{\pi}{L}f_{\Delta P}.
\end{equation}
It represents that the variational frequency of period spacings $f_{\Delta P}$ is the $L/\pi$ times of the width of the $\mu$-gradient region in buoyancy size $\Lambda_{\mu}$. In addition, Equation \eqref{eq:Lfp} can be expressed as the form of
\begin{equation}\label{eq:Lfp1}
f_{\Delta P}^{-1}=\frac{\pi}{L}\Lambda_{\mu}^{-1}.
\end{equation}
Its form is well equivalent to Equation \eqref{eq:DP-L0}. They (Equations \eqref{eq:DP-L0} and \eqref{eq:Lfp1}) indicate that the pattern of the oscillation periods of g-modes is the results of the superposition of two periodic functions if the stars are composed with a convective core and a radiative envelope (sometimes with one or more thin convective shells and/or a shallow convective envelope), such as SPB and $\gamma$ Dor stars \citep[also refer to e.g.,][]{Miglio2008MNRAS}. The quasi-uniform period spacing corresponds to the buoyancy radius $\Lambda_{0}$ (see Equation \eqref{eq:DP-L0}) which is decided by the whole star. However, the periodic variations of period spacings corresponds to the buoyancy size of $\mu$-gradient region, i.e., $\Lambda_{\mu}$ (see Equation \eqref{eq:Lfp1}).

For dipole modes, i.e., $l=1$, and then $\Lambda_{\mu}=\pi/\sqrt{2}f_{\Delta P}\simeq2.22f_{\Delta P}$, which is shown with black dashed-line in Figure \ref{fig.2.2}. We directly fit all of those models with a linear function ($\Lambda_{\mu}=a*f_{\Delta P}+b$) by the means of {\bf\small gnuplot}\footnote{gnuplot homepage: http://www.gnuplot.info/}  software and finally obtain
\begin{equation}
\Lambda_{\mu}=(2.04\pm0.01)f_{\Delta P}+(0.62\pm0.05)\times10^{-5} {\rm Hz}.
\end{equation}
It is shown with blue dash-line in Figure \ref{fig.2.2}. The relative uncertainties of the two fitting coefficients are 0.6\% and 8.8\%, respectively. The fitting result is very close to the relation of $\Lambda_{\mu}=2f_{\Delta P}$ which is presented with green dashed-line in Figure \ref{fig.2.2}.

As shown in Figure \ref{fig.2.2} all of the models follow the fitting relation, except for those models which have larger age and larger extra diffusion coefficient (presented with purple filled circles). In addition, those younger stars which just enter the main-sequence evolutionary stage and begin the hydrogen burning in stellar center also depart from the fitting relation. They have smaller $f_{\Delta P}$.

In Figure \ref{fig.2.2}, the four black points represent HD 50230, KIC 8324482, KIC 10526594, and KIC 10526594$^a$, respectively. Their $f_{\Delta P}$ are listed in Table \ref{table_1}. The $\Lambda_{\mu}$ is calculated from their best fitting models which is determined from asteroseismic analysis. The best fitting model of KIC 8324482 and HD 50230 are from \citet[][]{Deng2018} and \citet[][]{Wu2018}, respectively. For KIC 10526294, it is calculated by ourself according to the fundamental parameters and the inputs of \citet[][]{Moravveji2015AA}. They are $M_{\rm init}=3.25~\rm M_{\odot}$, $Z_{\rm init}=0.014$, $X_{\rm init}=0.71$, $f_{\rm ov}=0.017$, and $\log D_{\rm mix}=1.75$, and the central hydrogen mass fraction $X_{\rm C}=0.63$. It can be seen from Figure \ref{fig.2.2} that the three (or four) ``observations" are consistent with the theoretical model expected relation. Since lacking seismic modelling in details, the another example in Table \ref{table_1} --- KIC 6462033 --- is not included in Figure \ref{fig.2.2}.

It can seen from Figures \ref{fig.2}(a)-(b) and \ref{fig.2.1}(a)-(b) that both of $\Lambda_{\mu}$ and $f_{\Delta P}$ are affected by the metallicity, stellar mass, overshooting, and the extra diffusion. The variational behaviors between them are almost the same. Therefore, as shown in Figure \ref{fig.2.2} almost all of those calculated models are regularly located on a straight line except for some extra issues (with larger age and larger extra diffusion parameters $\log D_{\rm mix}$ especially for late stage of 3 and 4 $\rm M_{\odot}$ stars with $\log D_{\rm mix}=3.0$) as shown in Figure \ref{fig.2.1}.

\begin{figure}
  \begin{center}
\includegraphics[scale=0.43,angle=-90]{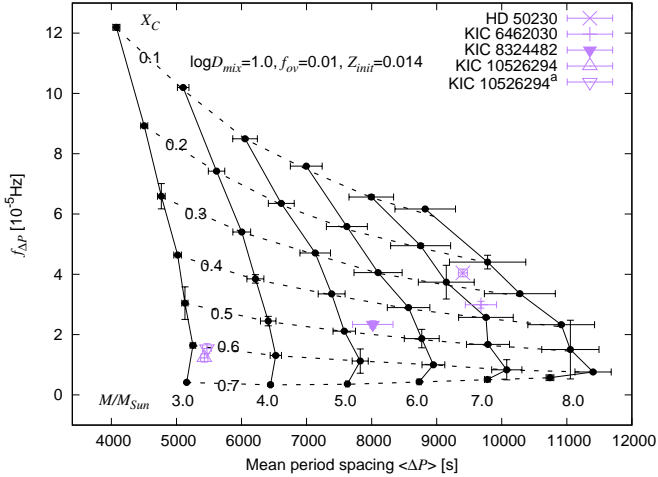}
   \caption{ $f_{\Delta P}$ vs. $\langle\Delta P\rangle$ (a new C-D-like diagram). Solid- and dished-lines represent the iso-mass lines and the contours of hydrogen mass fraction in stellar center ($X_{\rm C}$), respectively. Masses from 3.0 to 8.0 $\rm M_{\odot}$ with a step of 1.0 $\rm M_{\odot}$ and center hydrogen mass fraction $X_{\rm C}$ from 0.7 to 0.1 with a step of -0.1, respectively. The points represent the observations. The observational periods of KIC 10526294 comes from \citet[][]{Papics2014AA}, KIC 8324482 and 6462033 from \citet[][]{Zhang2018ApJ}, and HD 50230 from \citet[][]{Degroote2012AA}, respectively.
  }\label{fig.4}
  \end{center}
\end{figure}

\subsection{A new C-D-like diagram for SPB stars: $f_{\Delta P}$ vs. $\langle\Delta P\rangle$}

\citet[][]{jcd1984srps.conf} has been established a $D_{0}$ vs. $\Delta\nu$ diagram (so-called C-D diagram) to estimate the evolutionary stages and the stellar mass for low-mass main-sequence stars ($M\lesssim1.6~\rm M_{\odot}$; p-mode oscillations or solar-like oscillations). Correspondingly, other modified C-D diagrams ($\delta\nu_{02}$ vs. $\Delta\nu$ and $\delta\nu_{02}/\Delta\nu$ vs. $\Delta\nu$ diagrams) are usually used in the research of solar-like oscillations \citep[see e.g.,][]{White2011ApJa,White2011ApJb}. That is based on the different dependencies of the large ($\Delta\nu$) and small ($D_{0}$ or $\delta\nu_{02}$) separations. $\Delta\nu$ is sensitive to the stellar mass (the whole star) and $D_{0}$ and/or $\delta\nu_{02}$ are sensitive to the stellar center, i.e., the evolutionary status \citep[see e.g.,][]{jcd1984srps.conf,Aerts2010}.

As shown in Figure \ref{fig.1}(a) the convective core reduces and leaves a wider and wider $\mu$-gradient region when the central hydrogen mass fraction $X_{\rm C}$ decreases. As a matter of fact, convective core fast increase on the early stage of the main sequence and then slowly decrease until it fully disappearing. Compared with convective core decreasing, the time of convective core increasing is so short and can be ignored. Obviously, the wider $\mu$-gradient region corresponds to larger buoyancy radius $\Lambda_{\mu}$ (see Figure \ref{fig.1}(b) and Equation \eqref{eq:Lambdamu}). In addition, according to the relation between $f_{\Delta P}$ and $\Lambda_{\mu}$ (shown in Figures \ref{fig.2}, \ref{fig.2.1}, and \ref{fig.2.2}), it can be easily concluded that $f_{\Delta P}$ is sensitive to the central status for SPB stars. In a word, $f_{\Delta P}$ can be therefore used as a director to probe the central hydrogen mass fraction $X_{\rm C}$, i.e., the evolutionary status.

On the other hand, as shown in Figures \ref{fig.2}(c) and \ref{fig.2.1}(c) the period spacing $\Delta P$ is sensitive to the stellar mass $M$ and increases with the increasing of $M$ \citep[also refer to e.g.,][]{Moravveji2015AA}. Surely, the period spacings $\Delta P$ is also sensitive to the evolutionary stages for a given evolutionary track. But their variational behaviors slightly differ from each other for different masses. In addition, as shown in Figures \ref{fig.2} and \ref{fig.2.1} the period spacing $\Delta P$ is also affected by the metallicity and the overshooting in convective core \citep[also refer to e.g.,][]{Miglio2008MNRAS}.

Based on the above analysis, a new C-D-like diagram is constructed ($f_{\Delta P}$ vs. $\langle\Delta P\rangle$ diagram; see Figure \ref{fig.4}) for the SPB stars. Similar to the C-D diagram of \citet[][]{jcd1984srps.conf}, the new C-D-like diagram also can be used to roughly estimate the stellar mass and constrain their evolutionary stages.

\subsubsection{Comparing with the other asteroseismic analysis results}
Four stars, KIC 10526294, 8324482 and 6462033, and HD 50230, are symbolled in Figures \ref{fig.4} and \ref{fig.5} with different point types. Three of them (KIC 10526294, KIC 8324482, and HD 50230) are modelled with asteroseismology analysis. The corresponding fundamental parameters are also determined.

It can be seen from Figure \ref{fig.4} that the central hydrogen mass fraction of KIC 10526294 is slightly larger than 0.60 and far less than 0.70 (to be around $0.60-0.65$ in Figure \ref{fig.4}). On the other hand, its mass is slightly larger than 3.0 $\rm M_{\odot}$ and far less than 4.0 $\rm M_{\odot}$. They are consistent with the results of \citet[][$X_{\rm C}\simeq0.63$, $M=3.25~\rm M_{\odot}$]{Moravveji2015AA}.

As shown in Figure \ref{fig.4} the central hydrogen mass fraction of HD 50230 is slightly smaller than 0.30. Its mass $M$ is near and slightly larger than 7.0 $\rm M_{\odot}$. In the work of \citet[][]{Degroote2010Natur}, they made brief seismic analysis and found that HD 50230 has a mass between 7 and 8 $\rm M_{\odot}$ and about 60\% of its initial hydrogen in the center has already been consumed. For solar mixture, 40\% of its initial hydrogen corresponds to the mass fraction of 0.28 which is fully consistent with our result. In our another work \citep[][]{Wu2018}, we seismically modelling it and find that the central hydrogen mass fraction of the best fitting model is $X_{\rm C}=0.306$ with a mass of $M\simeq6.2~{\rm M_{\odot}}$. The evolutionary status is consistent with the present work predicted. For stellar mass, the predicted mass in the present work is far larger than that of \citet[][]{Wu2018} and consistent with that of \citet[][]{Degroote2010Natur}. The discrepancies among them are leaded by the other input physics which affect the final best fitting model (see the next section).

For KIC 8324482, the asteroseismic analysis \citep[][]{Deng2018} shows that its mass is around $M=4.75~{\rm M_{\odot}}$. Correspondingly, the mass fraction of central hydrogen is $X_{\rm C}=0.48-0.49$. They are near to the C-D-like diagram (Figure \ref{fig.4}) predicted: mass is near to 5.0 $\rm M_{\odot}$ and $X_{\rm C}$ to around 0.5.

\subsubsection{Effect factors for the new C-D-like diagram}

\begin{figure*}
  \begin{center}
\includegraphics[scale=0.50,angle=-90]{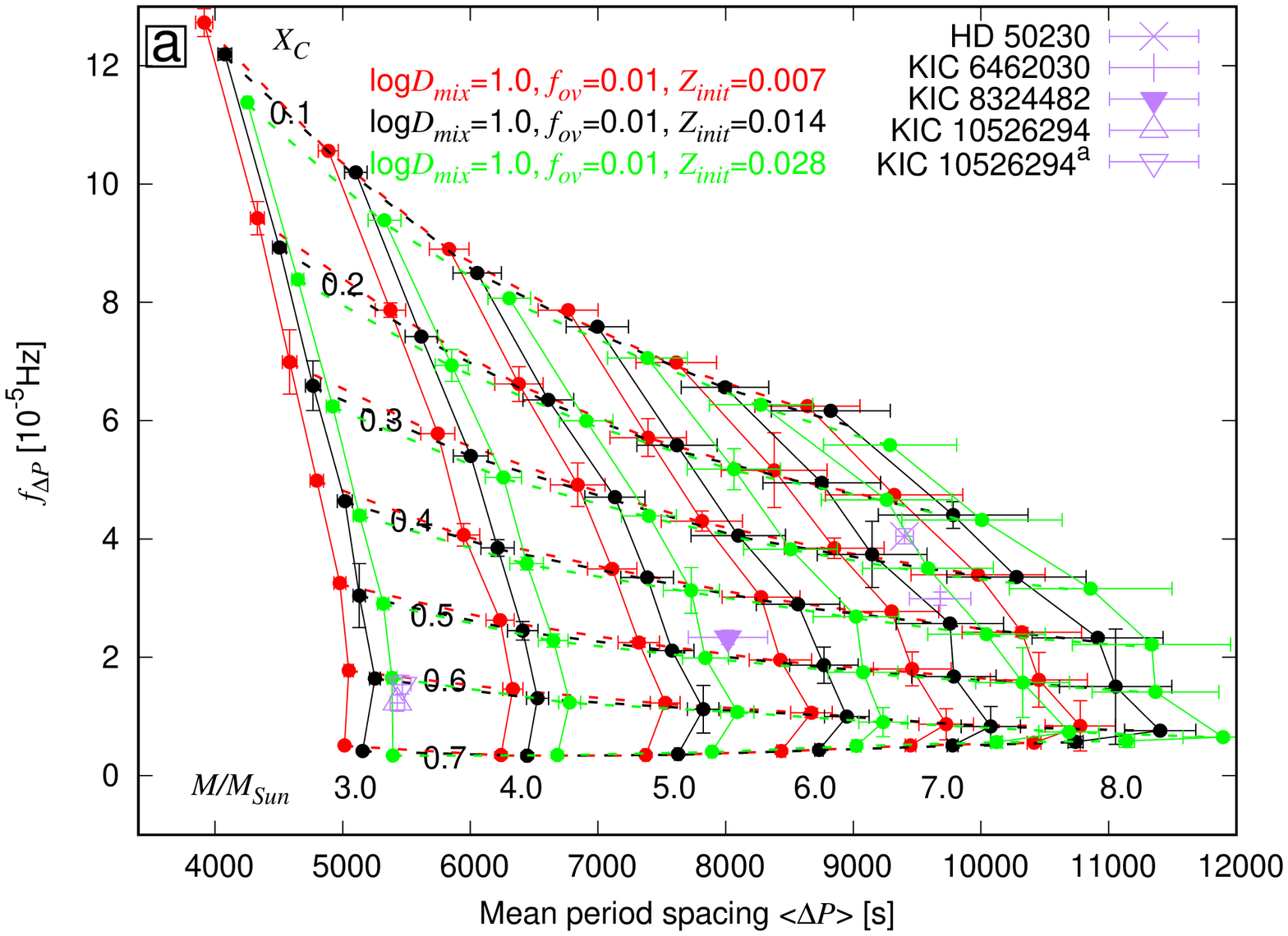}
\includegraphics[scale=0.50,angle=-90]{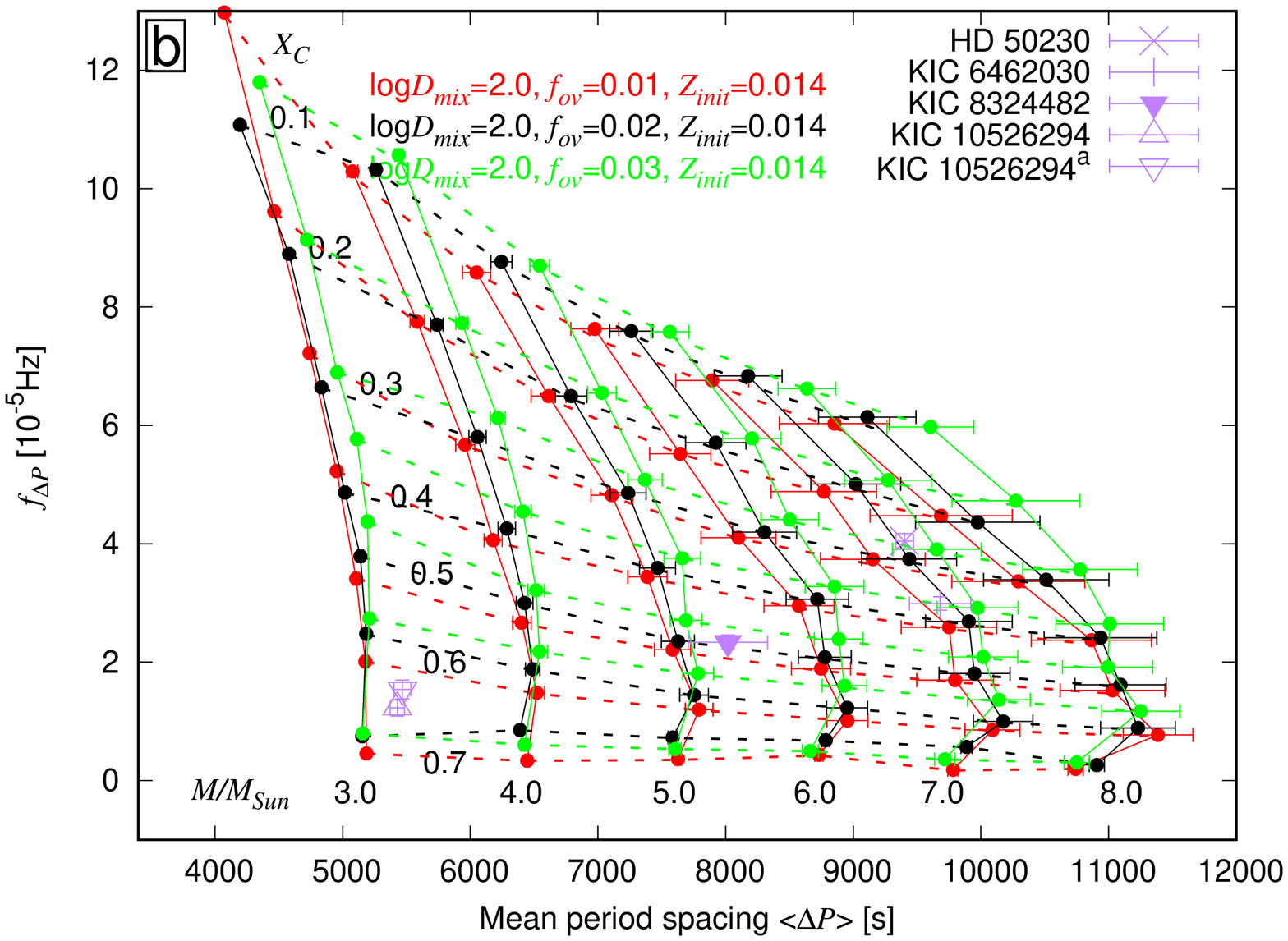}
\includegraphics[scale=0.50,angle=-90]{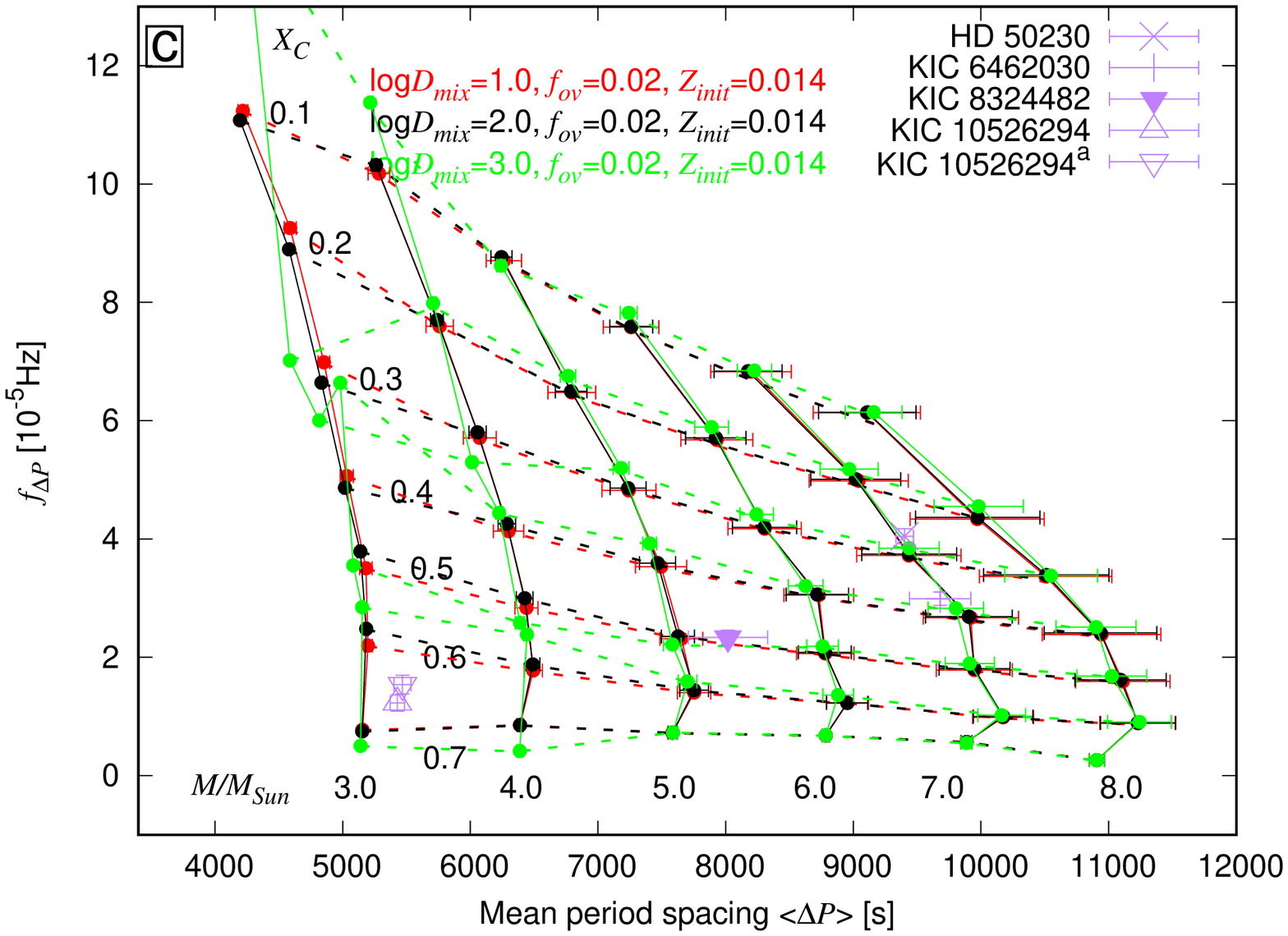}
   \caption{Similar to Figure \ref{fig.4}, but for multi-metallicites, multi-overshooting, and multi-extra diffusion in upper (panel (a)), middle (panel (b)), and bottom (panel (c)) panels, respectively. In panel (c), the irregular variation for green grid with mass $M=3.0~{\rm M_{\odot}}$ corresponds to irregular green lines in panel (b) of Figure \ref{fig.2.2} which have larger extra-mixing ($\log D_{\rm mix}=3.0$).
  }\label{fig.5}
  \end{center}
\end{figure*}

\begin{figure}
  \begin{center}
\includegraphics[scale=0.4,angle=-90]{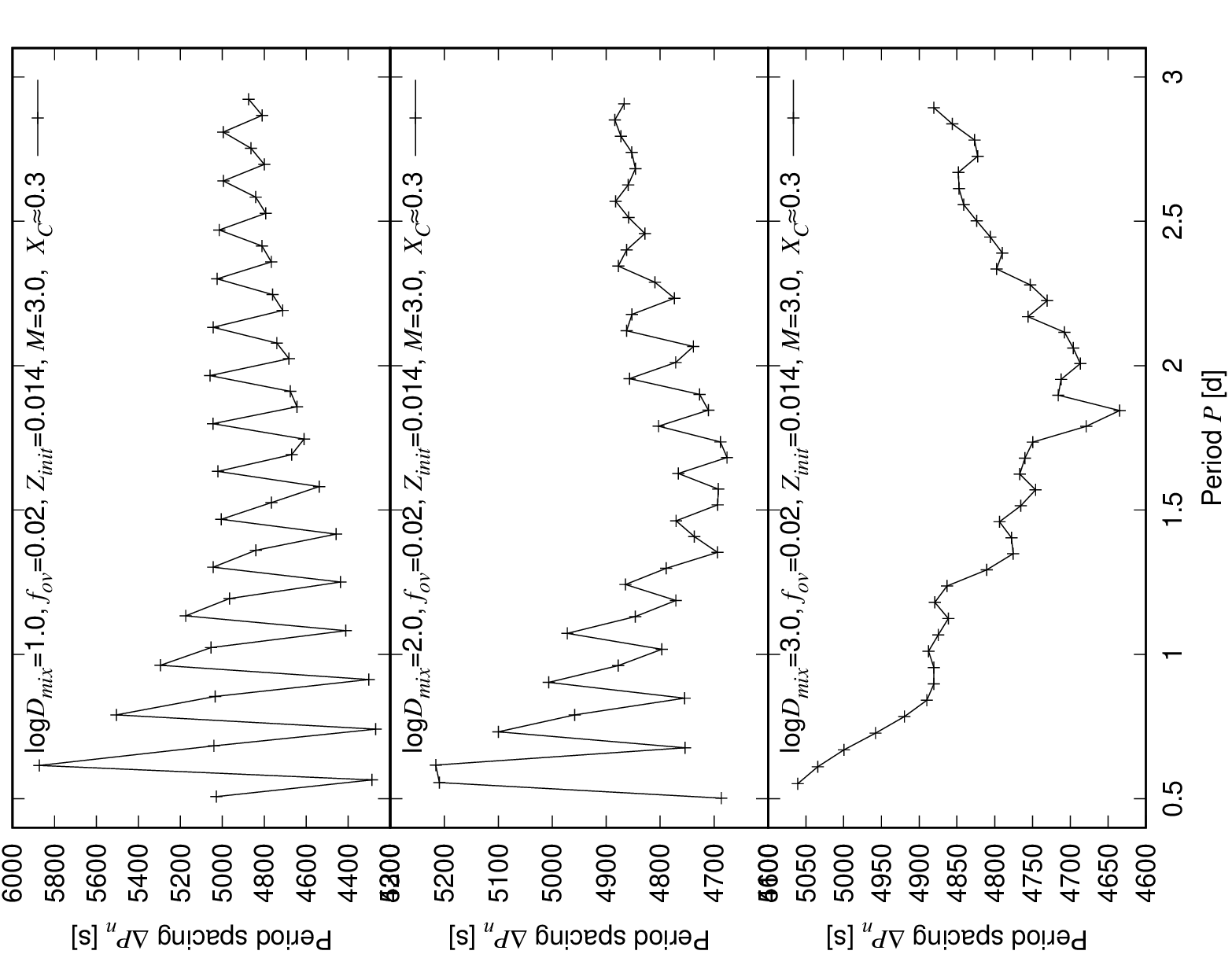}
   \caption{Period spacings $\Delta P_{n}$ vs. Periods $P$ of $l=1$, $m=0$ modes for three models. Their masses, overshooting parameters in convective core, the initial metal mass fraction, and the evolutionary status are the same: $M=3.0~{\rm M_{\odot}}$, $f_{\rm ov}=0.020$, $Z_{\rm init}=0.014$, and $X_{\rm C}\simeq0.3$. Their extra diffusion coefficients $\log D_{\rm mix}$ are different among the three models. They are 1.0, 2.0, and 3.0, for upper, middle, and bottom panels, respectively.
  }\label{fig.6}
  \end{center}
\end{figure}

According to the theory of stellar structure and evolutions, many factors will affect the width of the $\mu$-gradient region and the shape of buoyancy frequency in SPB stars, such as rotation, overshooting, extra-mixing \citep[see e.g.,][]{Miglio2008MNRAS,Moravveji2015AA}. In addition, the structure of stellar model are mainly dependent on the initial element composition due to the inner hydrogen-burning and matter opacities.

In the present work, we will discuss the effects which come from different metal mass fraction, overshooting, and extra-mixing. They are shown in Figure \ref{fig.5}. As shown in upper panel of Figure \ref{fig.5} except the initial metal mass fractions are different the other inputs are them same ($\log D_{\rm mix}=1.0$ and $f_{\rm ov}=0.010$). The initial metal mass fractions are $Z_{\rm init}=0.007$ (poor metal; red), $0.014$ (near-solar; black), and $0.028$ (richer-metal; green), respectively. It can be found from the upper panel of Figure \ref{fig.5} that a higher metal mass fraction $Z_{\rm init}$ leads to a larger period spacings $\langle\Delta P\rangle$ at a given evolutionary stage if the other initial inputs are the same (also see Figure \ref{fig.2}(c)). Correspondingly, they have smaller variational frequency on period spacings ($f_{\Delta P}$; see Figures \ref{fig.5} and \ref{fig.2}(a)).

It can be seen from Figure \ref{fig.5} that the increasing or decreasing of $Z_{\rm init}$ are equivalent to be of slightly increasing or decreasing stellar mass $M$ on the new C-D-like diagram ($f_{\Delta P}$ vs. $\langle\Delta P\rangle$ diagram). The influences slightly increase with the increasing of stellar mass. It seems to move the C-D-like diagram as a whole along with the iso-$X_{\rm C}$ line. The influence on $f_{\Delta P}$ slightly increase with stellar evolving especially for those lower-mass stars ($M=3,~4~{\rm M_{\odot}}$). In a word, the new C-D-like diagram predicted $X_{\rm C}$ is almost independent of the metal mass fractions $Z_{\rm init}$ (the slight effects can be ignored in here). The predicted stellar mass $M$ will decrease with the increasing of $Z_{\rm init}$. For instance, the predicted masses of KIC 6462030 are around 7.5, 7.2, and 6.8 $\rm M_{\odot}$ for the grids of $Z_{\rm init}=0.007$, 0.014, and 0.028, respectively.

The middle panel of Figure \ref{fig.5} represents the differences of grids among different overshooting extensions ($f_{\rm ov}=0.010$, 0.020, and 0.030). The other inputs are also the same ($\log D_{\rm mix}=2.0$ and $Z_{\rm init}=0.014$). It can be seen from this panel that $f_{\Delta P}$ almost increase with the increasing of overshooting $f_{\rm ov}$. The influence merely slightly change for different evolutionary status. On the other hands, the mean period spacings $\langle\Delta P\rangle$ are seriously affected by the differences of overshooting. The influences change with stellar evolution. The differences between the different overshooting is very small which almost can be ignored on the early evolutionary stage and very large on the middle and late evolutionary stages. For predicting the stellar mass and evolutionary status, lower overshooting C-D-like diagram might give a larger mass and slightly older (i.e., smaller $X_{\rm C}$) star. For instance, in middle panel of Figure \ref{fig.5}, the stellar masses of HD 50230 are about 7.5, 7.0, and 6.8 $\rm M_{\odot}$ for $f_{\rm ov}=0.010$, 0.020, and 0.030, respectively.
The corresponding central hydrogen mass fractions $X_{\rm C}$ are around 0.25, 0.27, and 0.30 for $f_{\rm ov}=0.010$, 0.020, and 0.030, respectively. This is because that the larger overshooting leads to larger period spacings \citep[see also e.g.,][]{Miglio2008MNRAS,Moravveji2015AA}. Especially for late evolutionary stage stars, the convective core becomes smaller and smaller. The weight of overshooting for the whole convective core will become larger and larger.

In addition, the influences of the extra-mixing $\log D_{\rm mix}$ are shown in the bottom panel of Figure \ref{fig.5}. The panel illustrates three grids with three differen extra-mixing coefficients: $\log D_{\rm mix}=1.0$, 2.0, and 3.0. The core overshooting and the metal mass fraction are the same among the three grids ($f_{\rm ov}=0.020$ and $Z_{\rm init}=0.014$). It can be seen from this panel that the former two grids ($\log D_{\rm mix}=1.0$ and 2.0) almost overlap (also see Figure \ref{fig.2.1}). For the third grid ($\log D_{\rm mix}=3.0$), it also almost overlap with the former two for the larger mass. However, for the lower mass models ($M=3.0,~4.0~{\rm M_{\odot}}$), they are different especially for the late stages. The larger extra-mixing seriously smooths the $\mu$-gradient region and makes these oscillation modes can not be regularly trapped by this region.

Figure \ref{fig.6} represents the period spacings of three different models whose masses are 3.0 $\rm M_{\odot}$. The other initial inputs are the same except their extra-mixing parameters. In addition, they almost stay at similar evolutionary stages ($X_{\rm C}\simeq0.3$). The extra-mixing gradually increases from the upper (1.0) to the bottom (3.0) panels. It can be seen from those panels that the larger extra-mixing breaks the regular pattern of period spacings. Especially for the largest case, the periodical variation of period spacings almost disappears. In other words, the signal becomes weaker and weaker. In addition, it brings a larger periodical signal.

As shown in Table \ref{tab.logD.age} the previous asteroseismic analyses indicate that the extra-mixing is related to the stellar mass and age. For instance, KIC 10526294 and KIC 7760680 have the same masses (3.25 $\rm M_{\odot}$) but KIC 10526294 has smaller age ($\tau_{\rm age, KIC 10526294}\simeq63$ Myr) compared to KIC 7760680 ($\tau_{\rm age, KIC 7760680}\simeq202$ Myr). The optimal extra-mixing $\log D_{\rm mix}=1.75$ of 10526294 is larger than that of KIC 7760680 ($\log D_{\rm mix}=0.75$). Surely, such difference might be partly affected by the different of rotation. However, compared with KIC 10526294, HD 50230 has larger mass and similar age. Correspondingly, its extra-mixing ($\log D_{\rm mix}=3.8$) is far larger than that of KIC 10526294 ($\log D_{\rm mix}=1.75$).

Therefore, the influences of extra-mixing for the new C-D-like diagram might be ignored for real stars. The larger extra-mixing in late stage of lower masses may not exist in real stars.

\begin{deluxetable}{lcccc}
\centering
\tablecaption{Summary of part seismically analyzed SPB stars.\label{tab.logD.age}}
\tablehead{
\colhead{ID} &  \colhead{Mass} & \colhead{Age} & \colhead{$\log D_{\rm mix}$} & \colhead{Refs.}\\
&  \colhead{[$\rm M_{\odot}$]} & \colhead{[Myr]} & \colhead{} & \colhead{}
}
\startdata
KIC 7760680  & 3.25 & 202  & 0.75  & \citet[][]{Moravveji2016ApJ} \\
KIC 10526294 & 3.25 & 63   & 1.75  & \citet[][]{Moravveji2015AA} \\
KIC 8324482  & 4.75 & 90   & 2.0   & \citet[][]{Deng2018}  \\
HD 50230     & 6.2125 & 62 & 3.8   & \citet[][]{Wu2018}
\enddata
\end{deluxetable}

\section{Conclusions and Discussions}\label{sec.con.dis}
The oscillation properties of the SPB stars are high-order, low-degree g-modes with almost quasi-equal period spacings which varies with the stellar mass and the evolutionary stages. The period spacing presents clear deviations from the uniformity one. The deviation periodically varies with period. Such variations would be used to constrain the shape variations of buoyancy frequency $N$ beyond the convective core \citep[see also e.g.,][]{Moravveji2015AA}. In the present work, we make a series theoretical model calculations to analyze the period spacing variations. The investigation can be briefly concluded as follows:

\rmnum{1}: Based on the theoretical calculations, we find that the variational frequency of the period spacings ($f_{\Delta P}$) is related to the width of the $\mu$-gradient region ($\Lambda_{\mu}$) with the relation of $\Lambda_{\mu}\sim2f_{\Delta P}$. All of those models perfectly follows this law, except for the very early evolutionary stages SPB stars, for instance, the inner consumed hydrogen is less than 5\% of the initial hydrogen, and those models which have larger age and larger extra-mixing parameters (see Figure \ref{fig.2.2}). It means that the value of $f_{\Delta P}$ points the width of the $\mu$-gradient region and also the central hydrogen $X_{\rm C}$.

\rmnum{2}: Based on the different dependencies of $f_{\Delta P}$ and $\langle\Delta P\rangle$, we construct a new C-D-like diagram for SPB stars. It can be used to roughly constrain the stellar evolutionary stages (i.e., the central hydrogen $X_{\rm C}$) and estimate the stellar mass ($M$).

The expected $X_{\rm C}$ from the new C-D-like diagram is almost independent of metallicity and extra-mixing, and slightly affected by the core overshooting extension. However, the expected $M$ can be affected by many physical processes, such as convective overshooting in the core, metal mass fraction.

Note that, in the present work, the new C-D-like diagram is only valid for non-rotation stars and/or ultra-slow rotators, since the effects of Coriolis force for the stellar oscillations are not considered in theoretical models. In addition, the new C-D-like diagram is seriously affected by the larger extra-mixing for the late evolutionary stage of low-mass stars ($M\leqslant4.0~{\rm M_{\odot}}$) which have larger age (see Figure \ref{fig.5}).

\acknowledgments
This work is co-sponsored by the NSFC of China (Grant Nos. 11333006, 11503076, 11773064, and 11521303), and by Yunnan Applied Basic Research Projects (Grant No. 2017B008). The authors express their sincere thanks to NASA and the Kepler team for allowing them to work with and analyze the Kepler data making this work possible and also gratefully acknowledge the computing time granted by the Yunnan Observatories, and provided on the facilities at the Yunnan Observatories Supercomputing Platform. The Kepler Mission is funded by NASA's Science Mission Directorate. The authors also express their sincere thanks to Dr. Q.S. Zhang and Dr. J. Su for their productive advices, to Dr. J. Su for his suggestions in extracting frequencies, to Dr. Patrick Lenz for his helps in using Period04 to automatically extract frequencies with batch processing. And finally, the authors are cordially grateful to an anonymous referee for instructive advice and productive suggestions to improve this paper.

\appendix

\section{inlist file of pulse in MESA (V6208)}
\&star\_job  ! HD49385 \\

      create\_pre\_main\_sequence\_model = .true.\\
      kappa\_file\_prefix = 'gs98'\\
      change\_initial\_net = .true.\\
      new\_net\_name = 'o18\_and\_ne22.net'\\

 / ! end of star\_job namelist\\

\&controls\\

    initial\_mass =      0.70D+01\\
    initial\_z    =      0.28D-01\\
    initial\_y    =      0.282D+00\\
    overshoot\_f\_above\_burn\_h =       0.01\\

      calculate\_Brunt\_N2 = .true.\\
      !use\_brunt\_dlnRho\_form = .true.\\
      use\_brunt\_gradmuX\_form = .true. \\
	  which\_atm\_option = 'Eddington\_grey' \\
      max\_years\_for\_timestep = 0.5d6\\
      varcontrol\_target = 1d-3\\
      mesh\_delta\_coeff = 0.4\\
      max\_allowed\_nz =30000  ! maximum number of grid points allowed\\
      max\_model\_number = 70000 ! negative means no maximum\\
      xa\_central\_lower\_limit\_species(1) = 'h1'\\
      xa\_central\_lower\_limit(1) = 0.05\\
      mixing\_length\_alpha = 2. \\
      set\_min\_D\_mix =.true.\\
      min\_D\_mix = 100.d0 ! D\_mix will be at least this large\\
      min\_center\_Ye\_for\_min\_D\_mix = 0.4 ! min\_D\_mix is only used when center Ye $>$= this\\
      dH\_div\_H\_limit\_min\_H = 2d-1\\
      dH\_div\_H\_limit = 0.0005d0\\
      dH\_div\_H\_hard\_limit = 1d-2\\

/ ! end of controls namelist\\

\section{Appendix FIgures}
\begin{figure*}
  \begin{center}
\includegraphics[scale=0.4,angle=-90]{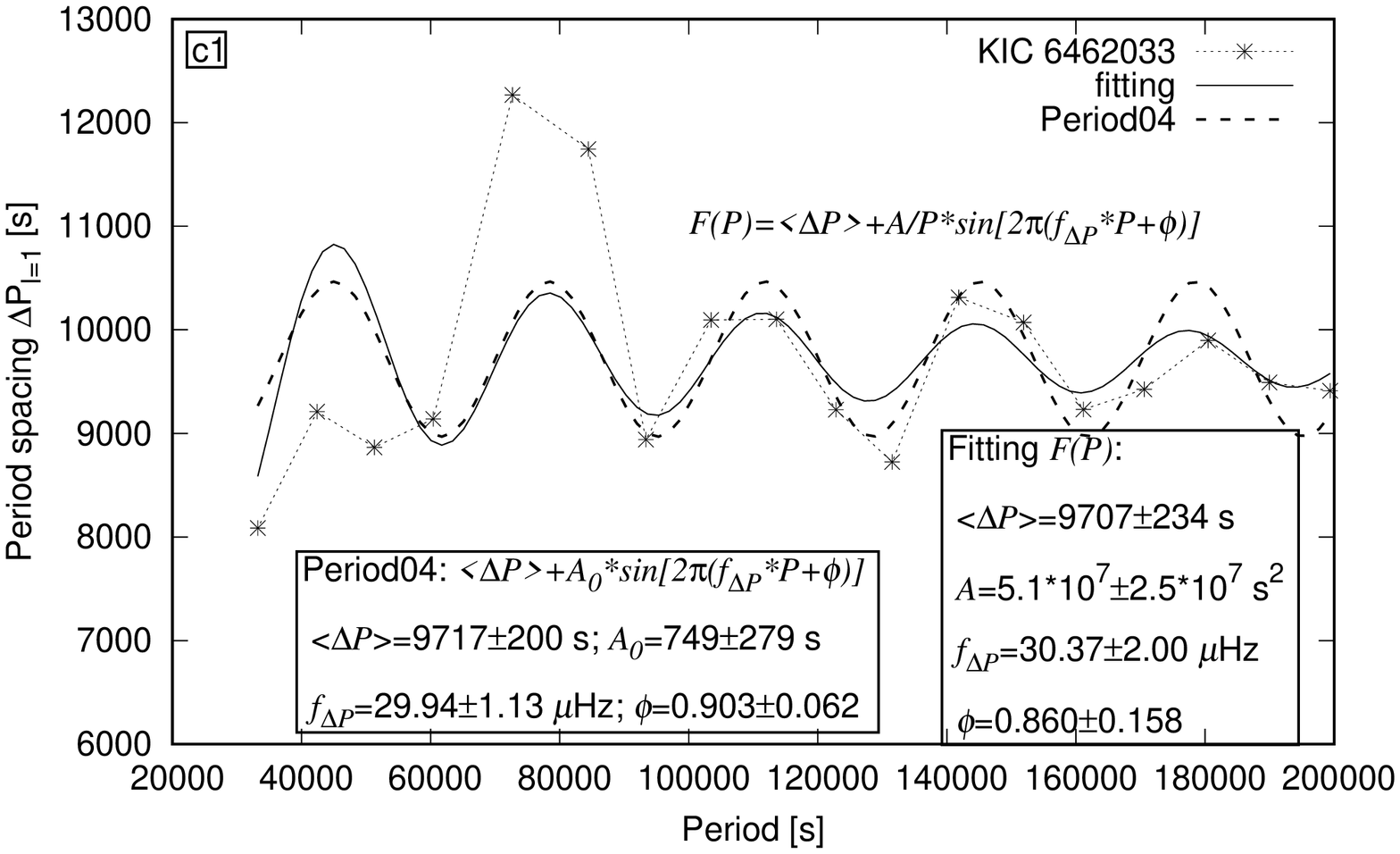}
\includegraphics[scale=0.4,angle=-90]{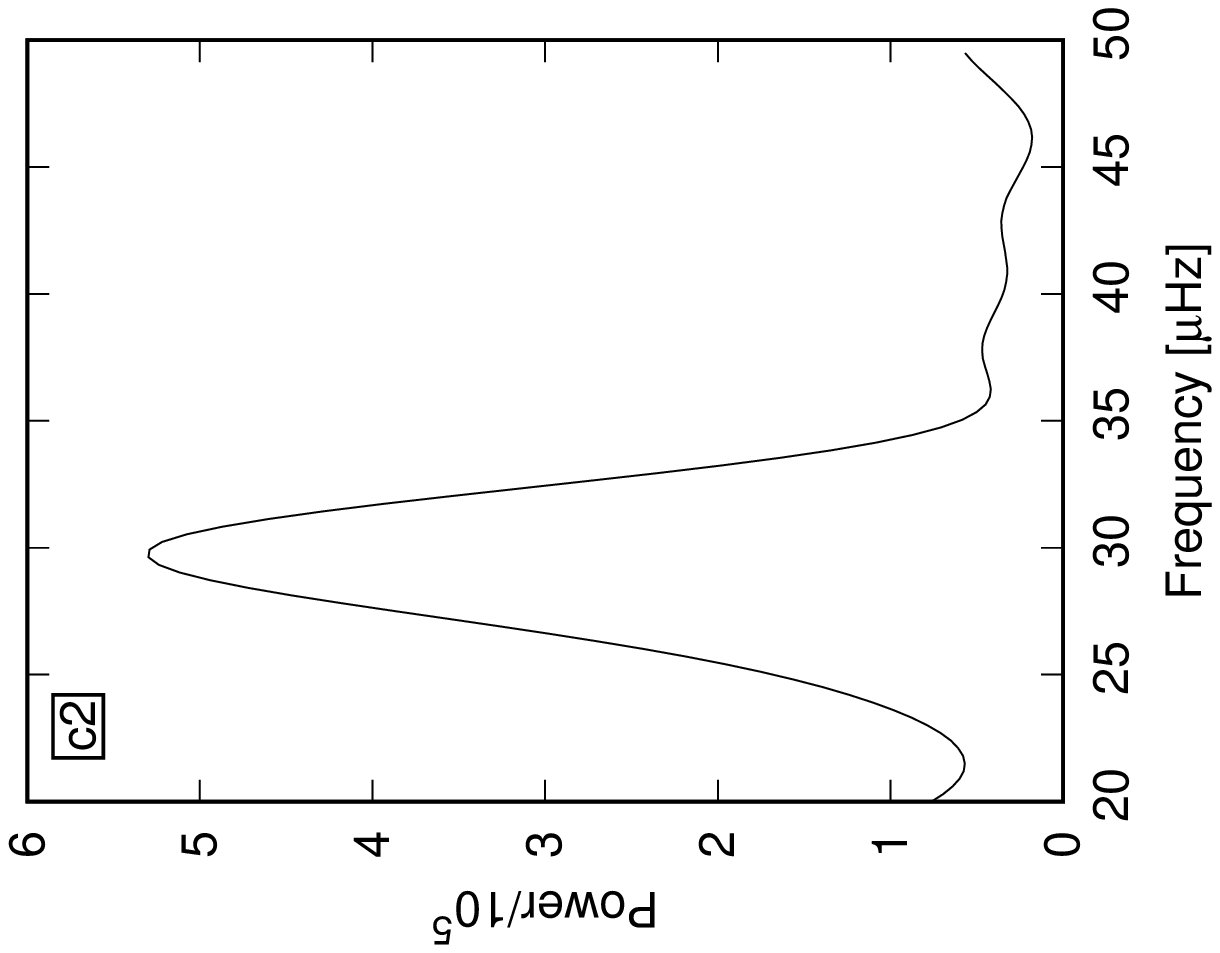}
\includegraphics[scale=0.4,angle=-90]{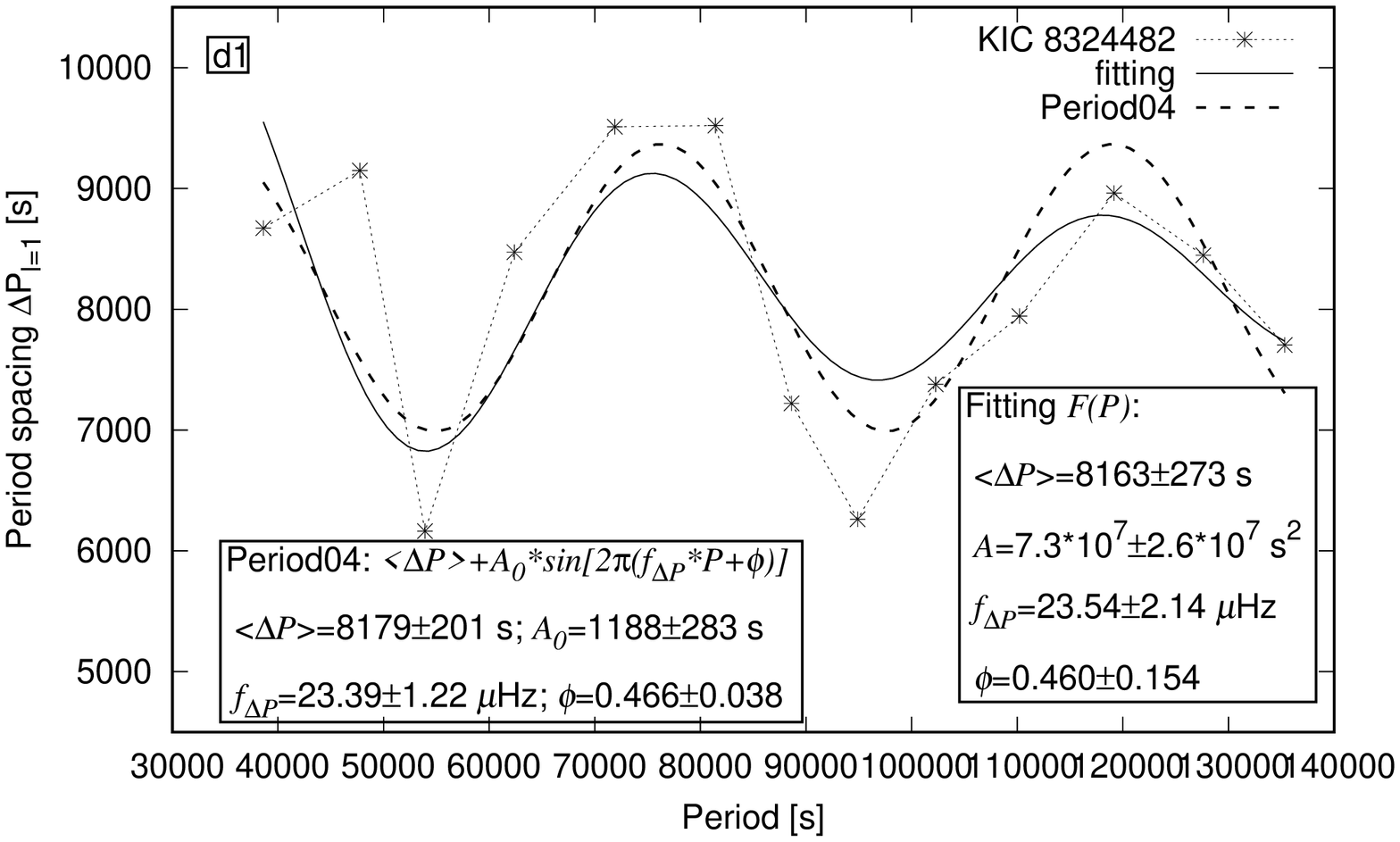}
\includegraphics[scale=0.4,angle=-90]{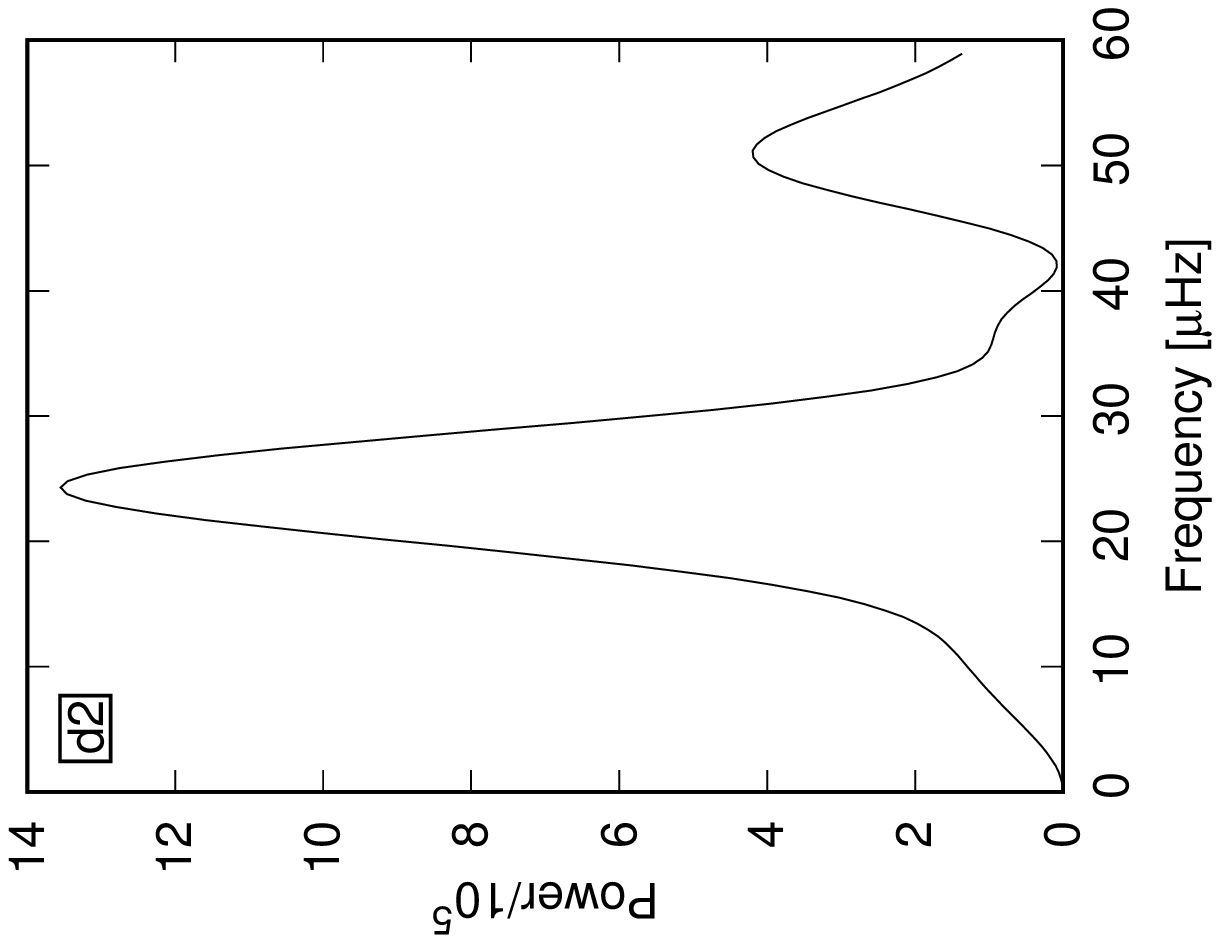}
\includegraphics[scale=0.4,angle=-90]{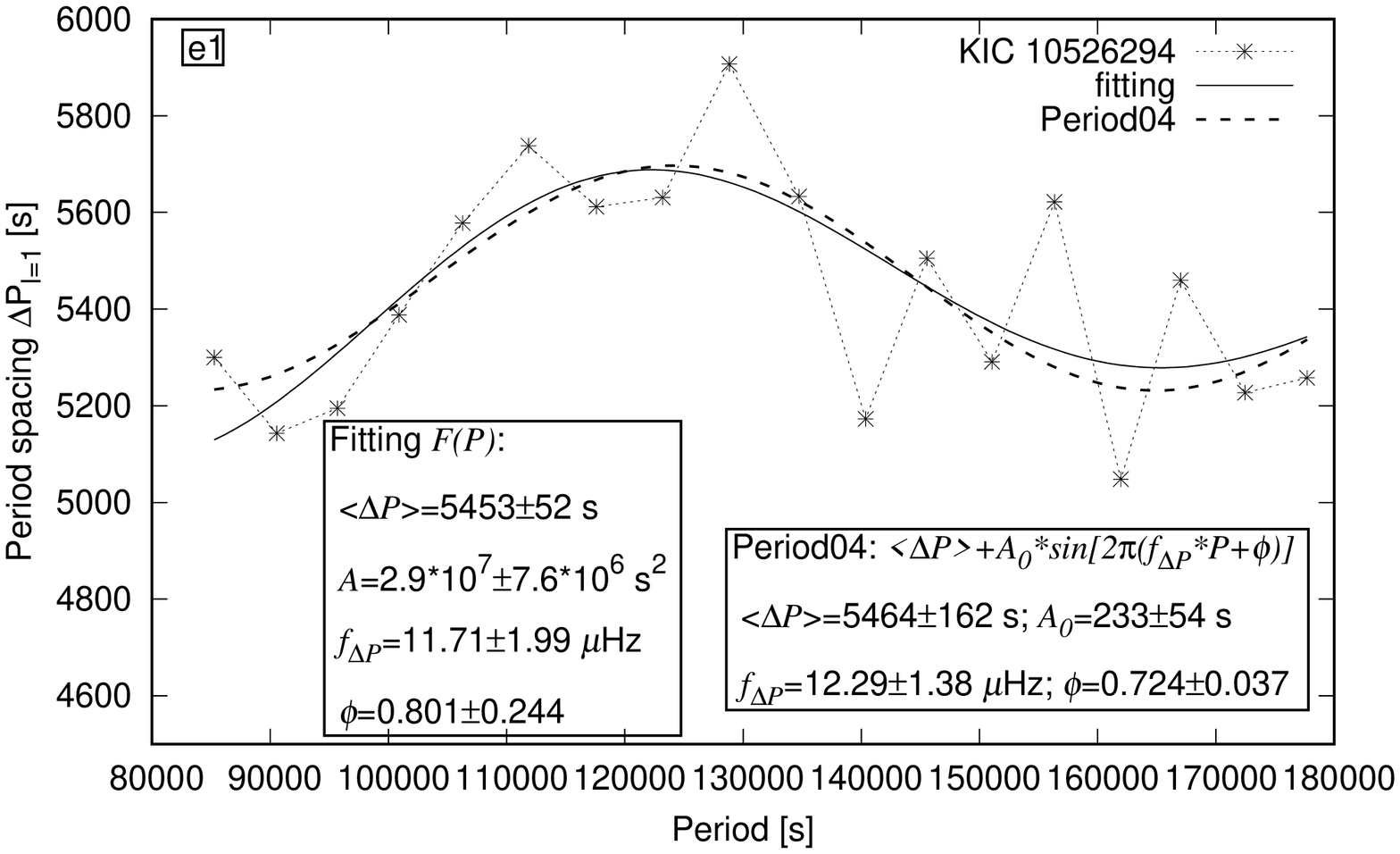}
\includegraphics[scale=0.4,angle=-90]{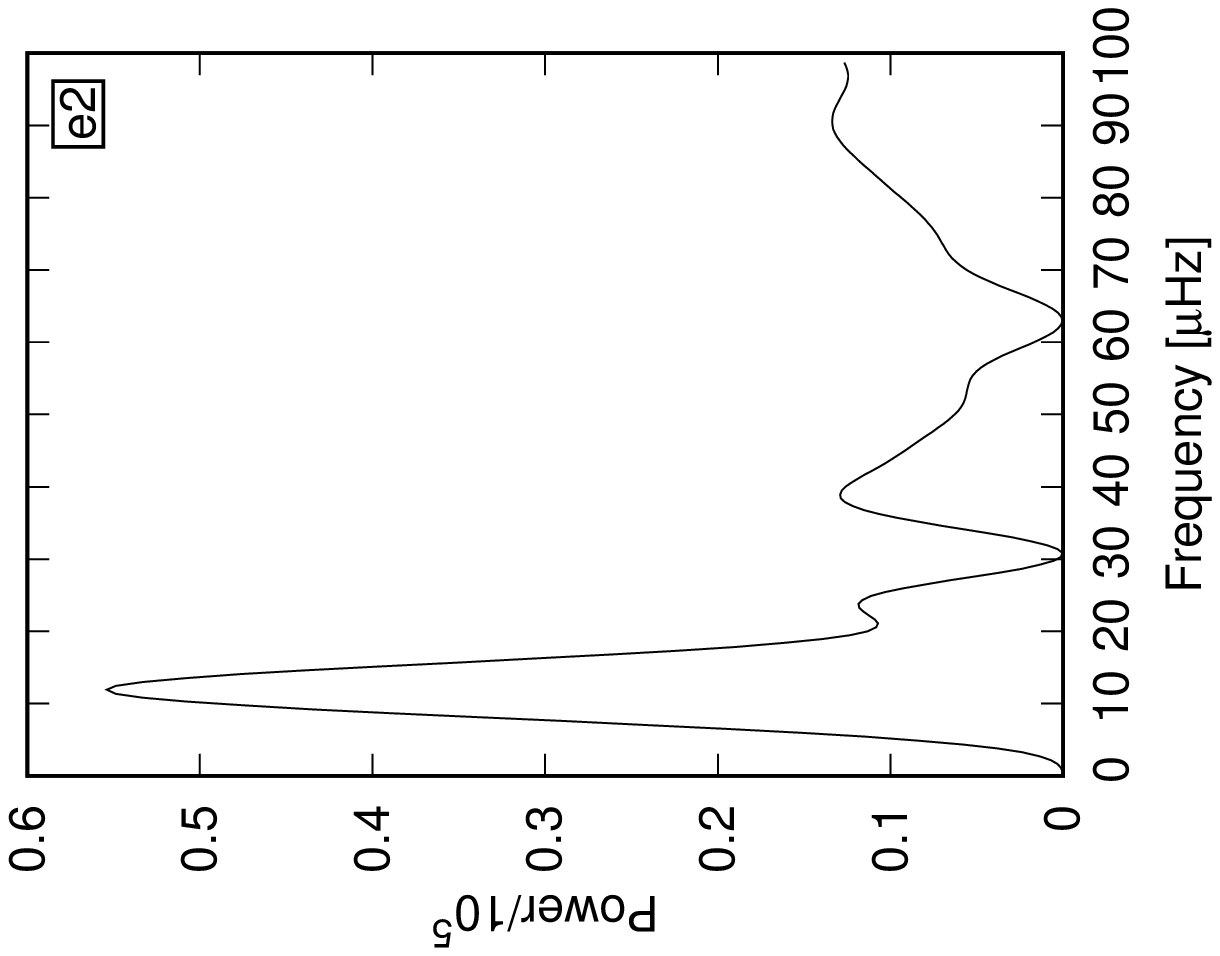}
\includegraphics[scale=0.4,angle=-90]{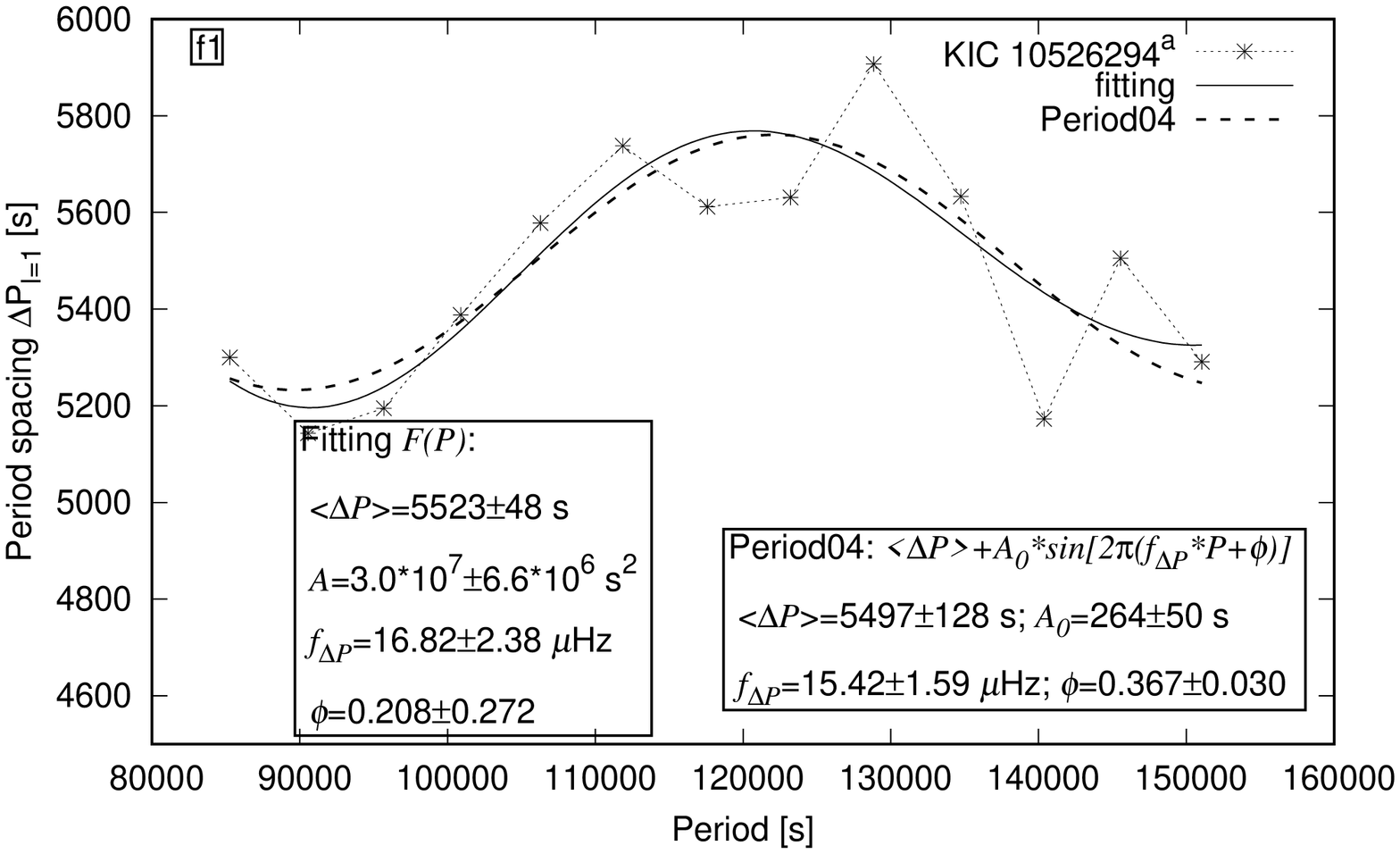}
\includegraphics[scale=0.4,angle=-90]{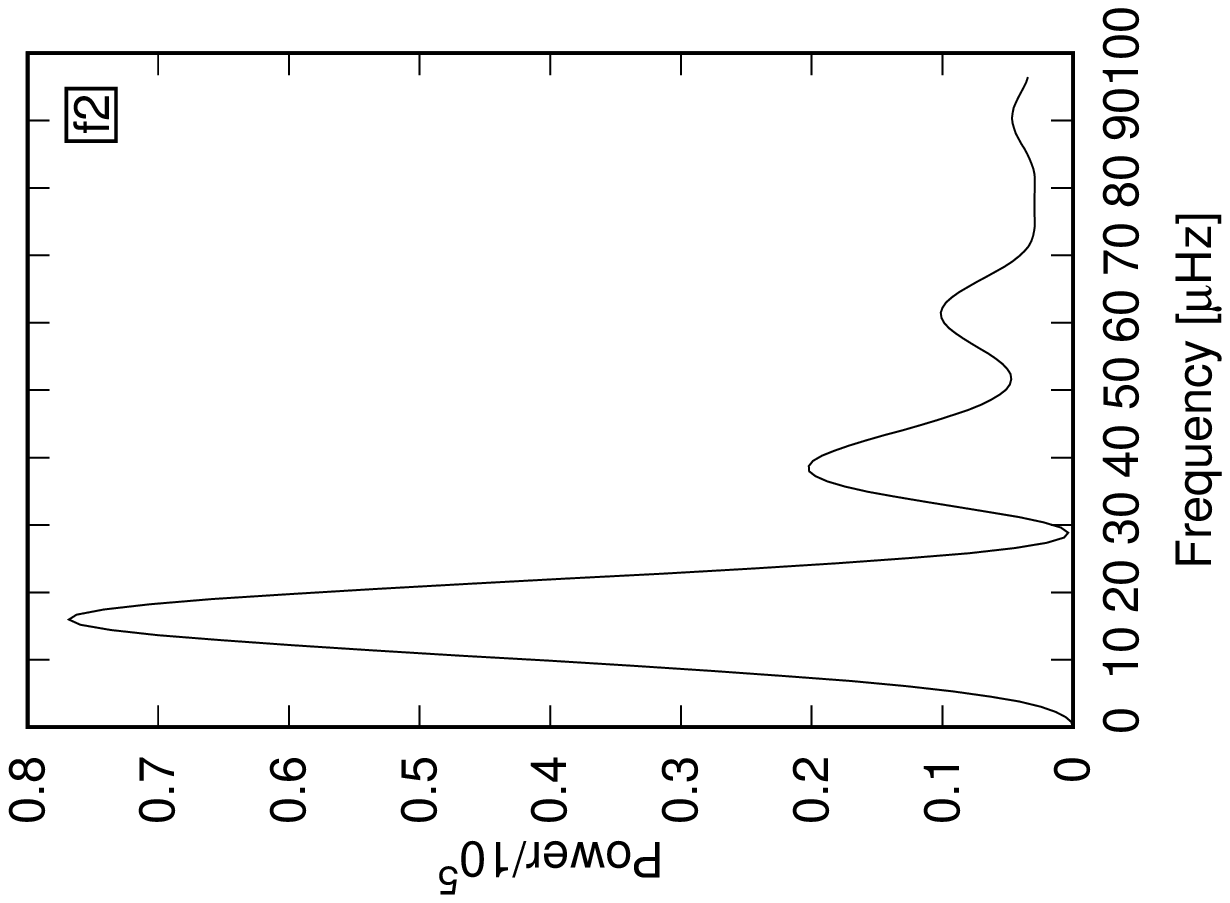}
   \caption{Similar to Panels b1 and b2 of Figure \ref{fig.fit1}, but for KIC 6462033 (panels c1 and c2), KIC 8324482 (panels d1 and d2), KIC 105262942 (panels e1 and e2), and KIC 105262942$^{\rm a}$ (panels f1 and f2), respectively.
  }\label{fig.A1}
  \end{center}
\end{figure*}





\end{document}